\documentclass[
 reprint,
 amsmath,
 pre,
 amssymb,
 nofootinbib,
 aps,
 longbibliography,
]{revtex4-1}

\usepackage[pdftex]{graphicx}
\usepackage{wasysym}
\usepackage{amsfonts}
\usepackage{ifsym}
\usepackage{pifont}
\usepackage{footmisc}

\makeatletter
\newlength{\apb@width}
\newcommand{\autoparbox}[2][c]{\settowidth{\apb@width}{#2}\parbox[#1]{\apb@width}{#2}}

\makeatother

\newcommand{\namedref}[2]{\hyperref[#2]{#1~\ref*{#2}}}

\newcommand{\tr}{\operatorname{tr}}

\newcommand{\pint}{\makebox[0pt][l]{\hspace{3.4pt}$-$}\int}

\renewcommand{\Re}{\mathop{\mathrm{Re}}}
\renewcommand{\Im}{\mathop{\mathrm{Im}}}

\newcommand{\Csphere}{{}^\bullet\kern-1.2pt C}
\newcommand{\Ctorus}{{}^\circ\kern-1.2pt C}

\newcommand{\nn}{\nonumber}

\newcommand{\COMMENT}[1]{}

\newcommand{\neqa}{\nonumber\end{eqnarray}}
\newcommand{\la}[1]{\label{#1}}

\newcommand{\<}{{\langle}}
\renewcommand{\>}{{\rangle}}

\newcommand{\re}{\relax{\rm I\kern-.18em R}}

\def\su2{{SU(2)}}

\def\[{\left[}
\def\]{\right]}

\def\({\left(}
\def\){\right)}
\def\[{\left[}
\def\]{\right]}

\def\<{\langle}
\def\>{\rangle}

\def\pint{-\hskip-0.41cm \int}
\def\i2{\frac{i}{2}}

\def\cO{{\cal O}}

\def\2F1{\,_2{\rm F}_1}

\usepackage[usenames,dvipsnames]{xcolor}
\usepackage{color}
\usepackage{graphicx}
\usepackage{dcolumn}
\usepackage{bm}
\usepackage{hyperref}

\usepackage{cancel}
\usepackage{ctable}
\usepackage{booktabs}
\newcolumntype{L}[1]{>{\raggedright\let\newline\\\arraybackslash\hspace{0pt}}m{#1}}
\newcolumntype{C}[1]{>{\centering\let\newline\\\arraybackslash\hspace{0pt}}m{#1}}
\newcolumntype{R}[1]{>{\raggedleft\let\newline\\\arraybackslash\hspace{0pt}}m{#1}}

\newcommand{\beq}{\begin{equation}}
\newcommand{\eeq}{\end{equation}}
\newcommand{\beqq}{\begin{equation*}}
\newcommand{\eeqq}{\end{equation*}}
\newcommand\beqa{\begin{eqnarray}}
\newcommand\eeqa{\end{eqnarray}}
\newcommand\beqaa{\begin{eqnarray*}}
\newcommand\eeqaa{\end{eqnarray*}}
\newcommand\bea{\begin{array}}
\newcommand\eea{\end{array}}

\begin{document}

\title{Rigorous bounds on the Analytic S-matrix}

\author{Andrea Guerrieri}
\author{Amit Sever}
\affiliation{School of Physics and Astronomy, Tel Aviv University, Ramat Aviv 69978, Israel}

\begin{abstract}
We consider a dual $S$-matrix Bootstrap approach in $d\geq 3$ space-time dimensions which relies solely on the rigorously proven analyticity, crossing, and unitarity properties of the scattering amplitudes. 
As a proof of principle, we provide rigorous upper and lower numerical bounds on the quartic coupling for the scattering of identical scalar particles in four dimensions.  
\end{abstract}

\pacs{Valid PACS appear here}
\maketitle

\section{Introduction}

An optimization problem can be viewed from either of two perspectives called respectively \emph{Primal} and \emph{Dual}.

In the revived non-perturbative $S$-matrix Bootstrap program of \cite{Paper2, Paper3}, the space of scattering amplitudes
is carved out numerically by solving an optimization problem in its \emph{Primal} form:
the target to optimize is a physical observable; the constraints are the physical principles of analyticity, crossing, and unitarity.
This approach has already been used to bound several classes of amplitudes. For example, in relation to integrable systems in two dimensions 
\cite{Doroud:2018szp,He:2018uxa,Cordova:2018uop,Paulos:2018fym,Homrich:2019cbt,EliasMiro:2019kyf,Cordova:2019lot,Bercini:2019vme,dualPaper,Kruczenski:2020ujw}, and in higher dimensions for the case of standard model physics \cite{Andrea,Bose:2020shm,Guerrieri:2020bto,Hebbar:2020ukp,Bose:2020cod} and quantum gravity theories \cite{Guerrieri:2021ivu}. At the same time, any bootstrap scheme in terms of the primal variables involves making an ansatz for the amplitude and some type of truncation: as a result, the bounds are not strictly rigorous.

Our main motivation for constructing a dual formulation stems from the
\emph{weak duality principle}. Suppose we maximize an observable $\mathcal{O}$ over some space $D$. Then for any value of the dual variables in the dual space $\bar D$, the dual function $\bar{\mathcal O}$ will always provide an upper bound on $\mathcal{O}$, independently of how hard the primal problem is
\beq
\mathcal{O}\le\(\mathcal{O}_* \equiv \max_D\, \mathcal{O} \)\  \leq\  \Big(\bar{\mathcal O}_*\equiv \min_{\bar D}\, \bar{\mathcal O}\, \Big)\le\bar{\mathcal O}\ .
\label{weakduality}
\eeq
This weak duality principle allows one to construct a bootstrap scheme for generating rigorous bounds on the $S$-matrix, thus placing it in a similar footing to the Conformal Bootstrap \cite{Poland:2018epd}. 
In this letter, we formulate a dual $S$-matrix Bootstrap problem in $d\geq 3$ for the scattering of identical scalar particles
that can be efficiently solved using SDPB \cite{Simmons-Duffin:2015qma, Landry:2019qug}.
A closely related -- albeit non-linear -- formulation was pioneered long ago in a series of papers  \cite{Lopez:1974cq,Lopez:1975wf,Lopez:1975ca,Bonnier:1975jz,Lopez:1976zs} and used to put rigorous bounds on the $\pi^0 \pi^0$ scattering amplitude in four dimensions. Moreover, an alternative dual formulation has been constructed recently using the Mandelstam representation \cite{He:2021eqn} -- see also \cite{Cordova:2019lot,dualPaper,Miro:2021rof} for previous examples in two dimensions. 

\begin{figure}[t]
\centering
\includegraphics[scale=0.162]{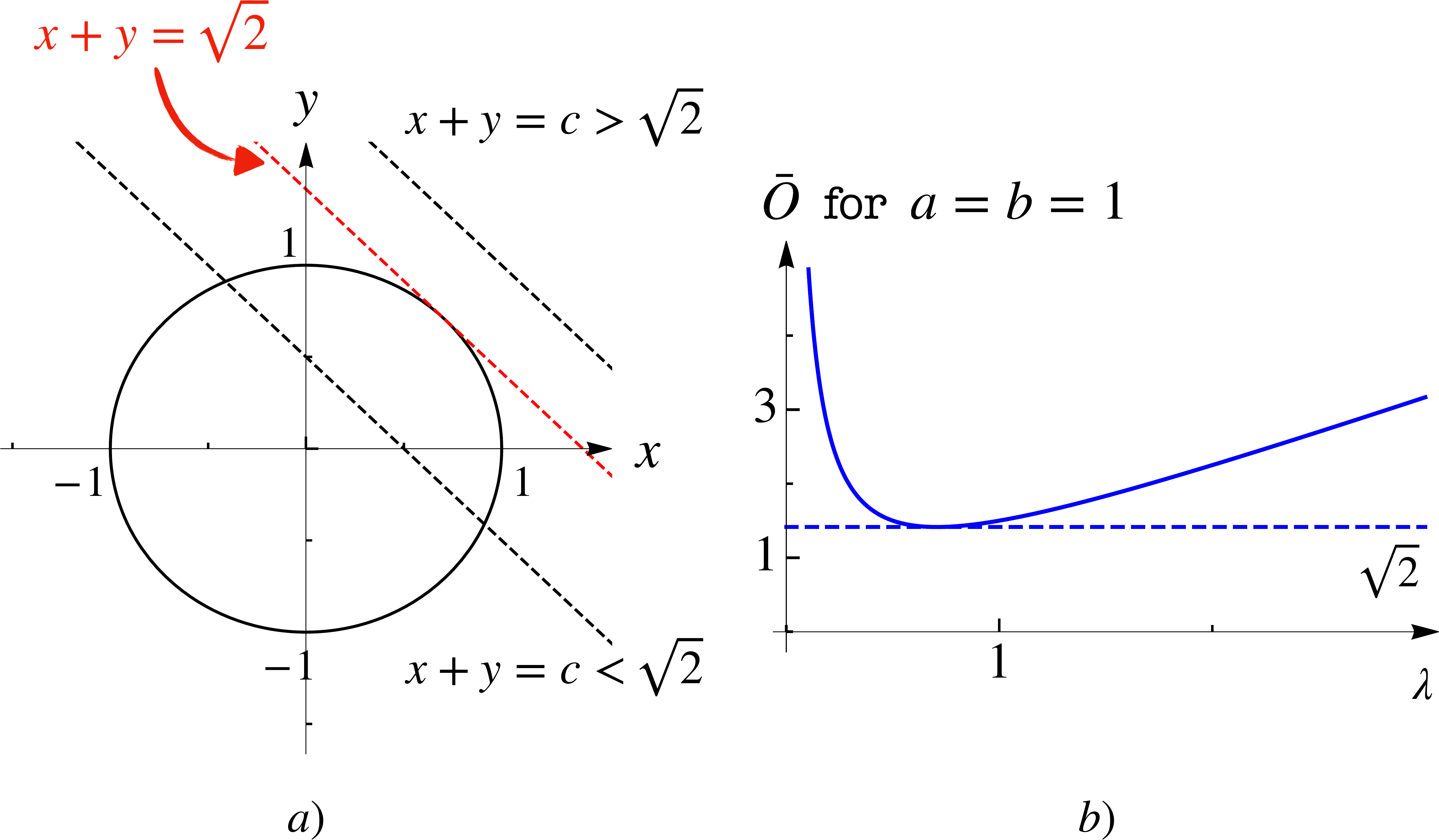}
\caption{\small a) The geometric solution of problem ${\rm max}\,\{x+y\,|\,x^2+y^2 \leq 1\}$. b) The dual function $\bar{\cal O}(\lambda)=\lambda+1/(2\lambda)$ of the same maximization problem. The minimum value is attained for $\lambda=1/\sqrt{2}$ and coincides with the optimal value of the maximization problem.}
\label{summary}
\end{figure}

Before illustrating our strategy, we shall review the logic behind the dual approach in a simpler example.\footnote{See Chapter 5 of \cite{cvx} for a general introduction to dual problems.}
We consider the following toy problem \footnote{You may think of the objective to maximize as the representative of a physical observable and the constraint a toy version of unitarity.}
\beq\label{toyproblem}
\mathcal{O}_*\equiv\underset{\{x,y\}}{\rm max}\,\{\cO(x,y)\ |\ x^2+y^2 \leq 1\}\ ,
\eeq
with the {\it primal objective} taken to be
\beq\la{Oab}
\cO(x,y)=a\,x+b\, y\ .
\eeq
The (real) variables $\{x,y\}$ are called \emph{primal variables}. 
The starting point to derive the dual version of problem \eqref{toyproblem} is to write the Lagrangian function 
\beq
\mathcal{L}(x,y|\lambda)\equiv(a\, x+b\, y)+\lambda(1-x^2-y^2)\ ,
\eeq
introducing a \emph{dual variable} for each constraint, as $\lambda\geq 0$.
This Lagrangian satisfies the following identity
\beq\label{ineq1}
\underset{\lambda}{\text{min  }}\{ \mathcal{L}|\lambda\ge0\}= 
\begin{cases}
    a\,x+ b\,y & x^2+y^2 \leq 1\\
    -\infty              & \text{otherwise}
\end{cases}\ .
\eeq
Combining this with the \emph{max-min inequality} yields
\begin{align}\label{maxmin}
\mathcal{O}_*\equiv&\underset{\{x,y\}}{\rm max}\,\Big\{\underset{\lambda}{\text{min }}\{ \mathcal{L}\,|\,\lambda\ge0\}\Big\} \\
\leq&\ \underset{\lambda}{\text{min  }}\,\Big\{\bar{\cO}(\lambda)\,|\,\lambda\ge0\Big\}\equiv\bar{\cal O}_*\ ,\nn
\end{align}
where the \emph{dual function} is defined as
\beq
\bar{\cO}(\lambda)\equiv\underset{\{x,y\}}{\text{max }} \mathcal{L}= \lambda+(a^2+b^2)/(4\lambda)\ .
\label{weakduality2}
\eeq

The last inequality in (\ref{maxmin}) proves the weak duality principle anticipated in (\ref{weakduality}).
\footnote{The case of a strong equality constraint, such as $(x^2+y^2:=1) =(x^2+y^2\le1)\wedge(x^2+y^2\ge1) $, can equally be formulated in this form using two dual variables or equivalently, a single standard unconstrained Lagrange multiplier.}

The dual objective in \eqref{weakduality2} is non-linear, but it is concave.\footnote{This is always true for the dual problem, even when the primal problem is non-convex.  It follows from the fact that the Lagrangian is an affine function of the dual variables and that the point-wise supremum operation preserves convexity. However, for non-convex problems, the duality gap does not necessarily close.}
For any fixed trial value of $\lambda\ge0$ the dual function provides a rigorous upper bound for the quantity $a\,x+b\,y$ of our toy problem.
The difference between primal and dual objective is called \emph{duality gap} and whether the duality gap is zero depends on the nature of the primal constraints.\footnote{{See \cite{vectoropt,mitopt,cvx} for a set of sufficient conditions.}}

In the following, we will apply this same logic to bound the quartic coupling between identical massive scalar particles in $d$-dimensions, showing explicit numerical results for $d=4$. 
We view our construction as a proof of principle that may be improved and optimized in the future.

\section{The quartic coupling problem}

We list the set of constraints for the $2\to2$ scattering amplitude of identical scalars. 
First, the analyticity of the amplitude in the cut $s$-plane can be imposed using dispersion relations. 
The amplitude satisfies fixed-$t$ double subtracted dispersion relations for any real $-{28}<t<4$ in the $s$-plane with two cuts starting at $s=4$ and $u=4$, all expressed in units of the mass, \cite{Martin:1965jj,Martin2,Roy:1971tc}. {Once combined with $s-u$ crossing, they can be put in the following form \cite{Roy:1971tc}} \footnote{See appendix \ref{disp_derivation} for a detailed derivation.} \beqa\label{dispersion}
&&0=\mathcal{A}(s,t)=T(s,t)-T(s_0,t_0)-\\
&&\frac{1}{\pi}\int\limits_{4}^\infty dv\, \[T_v(v,t) K(v,s,t;t_0){+} T_v (v,t_0) K(v,t,t_0;s_0)\]\,,\nonumber
\eeqa
where $T_v(v,t)= \text{Disc}_v T(v,t)$, and $(s_0,t_0)$ is an arbitrary subtraction point. Here, 
\beq\la{kernel}
K(v,s,t;t_0){=}\frac{1}{v{-}s}+\frac{1}{v{-}4{+}s{+}t}-\frac{1}{v{-}t_0}-\frac{1}{v{-}4{+}t{+}t_0}\,.
\eeq

A function that satisfies the $\cal A$ constraint is automatically analytic in the cut $s$-plane and $s-u$ crossing symmetric, but not necessarily $s-t$ symmetric. Hence, this equation must be supplemented by the crossing constraint, $T(s,t)=T(s,4-s-t)$. This implies that the primal variables must consist of $s-t$ crossing symmetric functions, which in turn supports only even spins.\footnote{Instead of imposing crossing as a constraint, one may use the manifestly crossing symmetric dispersion relations of \cite{Auberson:1973pf,Mahoux:1974ej}. 
We leave this possibility to a future exploration \cite{work_in_progress}.}

Unitarity is {most simply expressed as} the probability conservation for fixed energy $s$ and spin $\ell$
\beq\la{unitarity}
1=\sum_{n} |S_{2\to n}(s,\ell)|^2 \geq  |S_{2\to 2}(s,\ell)|^2\,,
\eeq
where $S_{2\to n}(s,\ell)$ denotes the matrix element between $2$-particle and $n$-particle states of spin $\ell$. The $S_{2\to 2}$ amplitudes are given by 
\beq
S_{2\to2} (s,\ell)=1+i \rho^2(s)f_\ell(s)\,,
\eeq
where $\rho^2(s)=\frac{(s-4)^{\frac{d-3}{2}}}{\sqrt{s}}$ is the $d$-dimensional two-particle phase space factor and $f_\ell(s)$ the partial wave projection\footnote{We work in the normalization of \cite{Correia:2020xtr} in which $P_\ell^{(d)}(z)= {}_2F_1\(-\ell,\ell+d-3,(d-2)/2,(1-z)/2\)$ and $\mathcal{N}_d=\frac{(16 \pi)^{(2-d)/2}}{\Gamma((d-2)/2)}$. In this normalization, the orthogonality of the partial waves takes the form ${{\cal N}_d\over2}\int_{-1}^1 d z\, (1-z^2)^{{d-4 \over 2}} P^{(d)}_\ell(z) P^{(d)}_{\tilde\ell}(z) =\delta_{\ell\tilde\ell}/ n_\ell^{(d)}$, with $n_\ell^{(d)}=\frac{(4\pi)^{d\over2}(d+2\ell-3) \Gamma (d+\ell-3)}{\pi\,\Gamma \((d-2)/2\) \Gamma (\ell+1)}$.}
\beq
f_\ell(s)=\frac{\mathcal{N}_d}{2}\int\limits_{-1}^1\! dz\, (1-z^2)^{\frac{d-4}{2}}P_\ell^{(d)}(z)T(s,t(z))\,,
\eeq
with $t(z)=\tfrac{1}{2}(s-4)(z-1)$. For later convenience, we 
re-phrase the unitarity constraint as the semi-definite (SDP) condition
\beq
\mathcal{U}_\ell(s>4)=\begin{pmatrix}
1-\frac{\rho^2}{2}\text{Im}f_\ell & \rho\, \text{Re} f_\ell \\
 \rho\, \text{Re} f_\ell & 2 \text{Im}f_\ell
\end{pmatrix}\succeq 0\,.
\label{sdpbunitarity}
\eeq

The equivalence between (\ref{unitarity}) and (\ref{sdpbunitarity}) can be seen by first noting that $\mathcal{U}_\ell$ is positive iff both its determinant and trace are positive. The positivity of the determinant is (\ref{unitarity}). The positivity of the trace for $s>4$ follows from that of the determinant, as can be seen by nothing that $\det \mathcal{U}_\ell=2\text{Im}f_\ell\tr \mathcal{U}_\ell-4\text{Im}f_\ell^2-\rho^2{\rm Re}f_\ell^2$.

Finally, in the elastic region of $4<s<16$, unitarity implies a stronger equality constraint
\beq\la{eu}
\det\mathcal{U}_\ell(16>s>4)=0\,,
\eeq
instead of the inequality (\ref{sdpbunitarity}).

The quantity we want to bound is the quartic coupling, defined as the value of the amplitude at the crossing symmetric point \cite{Paper3}
\beq\la{coupling}
g_0\equiv\frac{1}{2 n_0^{(d)}}T\left(\tfrac{4}{3},\tfrac{4}{3}\right)\,.
\eeq

Combining all the constraints, we write the Lagrangian \footnote{The sign $\pm$ depends on whether we want to maximize or minimize $g_0$.}
\beqa\label{Lagrangian}
\mathcal{L}^{\pm}&=&g_0{+}\iint_{\mathcal{D}}\!ds\,dt\,\mathcal{W}(s,t)\mathcal{A}(s,t)\\
&{+}&\sum_{\ell\,\text{even}}\bigg[\int\limits_4^{16}\!ds\,E_\ell(s)\det\mathcal{U}_\ell(s)\pm \int\limits_{4}^{\infty} ds \tr(\mathbf{\Lambda}_\ell(s)\mathcal{U}_\ell(s))\bigg]\,,\nn
\eeqa
where $\mathcal{W}(s,t)$ and $E_\ell(s)$ are unconstrained dual variables imposing 
the analyticity \eqref{dispersion} and elastic unitarity \eqref{eu} 
in their 
validity domains, $\cal D$ and $s\in [4,16]$. Here, crossing has already been solved at the level of the primal variables by restricting to even spins only, $f_{2n+1}(s)=0$. Finally, $\mathbf{\Lambda}_\ell(s)$ is a semidefinite positive matrix associated to the unitarity inequality constraint \eqref{sdpbunitarity} that we impose for all energies.\footnote{A simple theorem states that the integrand $\tr(\mathbf{\Lambda}_\ell(s)\mathcal{U}_\ell(s))$ is positive iff $\mathcal{U}_\ell(s) \succeq 0$ for any $\mathbf{\Lambda}_\ell(s)\succeq 0$.}

\subsection{The dual variable space}

Here comes the important advantage of the dual formulation: omitting part of the constraints may weaker the bound, but due to the inequality ${\rm min}_\lambda\{\bar{\cO}(\lambda)\}\le\bar{\cO}(0)$, it does not affect its rigor.

First, we simply set the dual variables associated with the elastic unitarity constraint (\ref{eu}) to zero, $E_\ell(s)=0$. This is because we do not know how to put it in SDP form. 
In the end, we find the elastic unitarity is satisfied in the region where the unitarity inequality is imposed -- see also sec. \ref{numerical_results}.
\footnote{It can be shown on general grounds that even with $E_\ell(s)=0$, the Lagrangian (\ref{Lagrangian}) leads to elastic functions, see \cite{Lopez:1976zs} and appendix \ref{phapp}. This is not in contradiction with the Ask theorem, \cite{Aks:1965qga}, because the S-matrix constraints are not imposed to an arbitrary large $s$.}

Next, we discuss the dual function $\mathcal{W}(s,t)$. Its analyticity properties are connected to those of its associated constraint $\mathcal{A}(s,t)$ and its domain of validity. 
We consider a sub-space of dual functions whose $s$-channel spin is bounded by a fixed integer $L$ 
\beqa\label{Wansatz}
\mathcal{W}_L(s,t(z))&=&\frac{\mathcal{N}_d}{2} (1{-}z^2)^{\frac{d-4}{2}}\sum_{\ell=0}^{L}w_{\ell}(s)\,P^{(d)}_{\ell}(z)\,,
\eeqa
where the sum runs over all spins, even and odd.
Note that even though we have solved the crossing constraint at the level of the primal variables by setting $f_{2n+1}(s)=0$, the odd spins dual variables $w_{2n+1}(s)$ impose non-trivial constraints. 
This is because the kernel in \eqref{dispersion} is not $s-t$ symmetric.
 
At this point, we shall specify the domain $\cal D$ of $\mathcal{W}_L$ where we impose the $\cal A$-constraint.
The larger this domain is, the more constraints we are imposing,  hence the stronger is the bound we get. The regime of validity of $\cal A$ is $-28<t<4$ and any $s$. 
In addition, we demand the integral of ${\cal W}(s,t)\,T(s,t)$ in (\ref{Lagrangian}) to be diagonal in spin. This is achieved by letting 
$z$ run over the range $z\in[-1,1]$. 
 
Combining the two conditions above implies that $0<s<32$. All in all, we take 
the integration domain to be $\mathcal{D}_\mu=\{(s,z)|z \in [-1,1]\land s \in[4,\mu^2]\}$, where $4<\mu^2<32$ is a free parameter that we will tune later when solving the dual problem numerically.  
Later we will also comment about the possibility of enlarging further this integration domain -- see Sec. \ref{dual_problem_section}.

With these choices, the $\cal A$-term in the Lagrangian (\ref{Lagrangian}) takes the form
\beqa\la{Wconstraint}
&&\iint_{\mathcal{D}_\mu}\!ds\,dt\, \mathcal{W}_L(s,t)\mathcal{A}(s,t)=\sum_{\ell=0}^{L} \int\limits_4^{\mu^2}ds\, w_{\ell}(s)a_{\ell}(s)\,,
\eeqa
where
\beqa
&&a_{\ell}(s)\equiv\frac{\mathcal{N}_d}{2}\int\limits_{-1}^1dz\, (1-z^2)^{\frac{d-4}{2}}P_{\ell}(z)\mathcal{A}(s,t(z))=\label{royequations}\\
&&\Re f_{\ell}(s)-\frac{\delta_{\ell,0}}{n_0^{(d)}}T(s_0,t_0)-\pint\limits_{4}^\infty dv \sum_{J \text{even}} k_{\ell,J}(v,s) \Im f_J(v),\nonumber
\eeqa
with the kernels $k_{\ell,J}$ given by
\beqa\la{kernels}
&&k_{\ell,J}(v,s)\equiv \frac{\mathcal{N}_d\, n_J^{(d)}}{2\pi} \int\limits_{-1}^1\!dz\, (1-z^2)^{\tfrac{d-4}{2}}P_\ell^{(d)}(z)\\
&&\times \[P_J(1{+}\tfrac{2t}{v-4}) K(v,s,t;t_0){+} P_J(1{+}\tfrac{2t_0}{v-4}) K(v,t,t_0;s_0)\]\,.\nn
\eeqa
In (\ref{royequations}) we used the fact that the imaginary part of the $\mathcal{A}$-constraint is automatically satisfied for real $s>4$. The even spin constraints $a_{2n}(s)=0$ are also known as \emph{Roy equations} \cite{Roy:1971tc}.
They relate the real part of each partial amplitude to the absorptive parts that can be measured experimentally and they have been successfully used 
in low energy QCD phenomenology, see \cite{Colangelo:2001df} for a review.
Notice also that the odd constraints $a_{2n+1}(s)=0$ do not contain the odd spin real parts as we set $\Re f_{2n+1}(s)=0$ by the choice of the primal variables.

\subsection{The dual problem}
\label{dual_problem_section}
By plugging (\ref{Wconstraint}) into (\ref{Lagrangian}), 
and choosing conveniently the subtraction point at $s_0{=}t_0{=}4/3$, the Lagrangian becomes
\beqa\label{finalLagrangian}
&&\mathcal{L}^\pm=g_0\Big(1-2\int\limits_4^{\mu^2} w_0(s)ds\Big)+\sum_{\ell=0}^{L} \int\limits_4^{\mu^2} ds\, w_{\ell}(s)\Re f_{\ell}(s)\\
&&{-}\int\limits_4^\infty dv \sum_{J \text{even}} \bar w_J(v)\Im f_J(v){\pm}\int\limits_4^\infty dv \sum_{J\,\text{even}} \tr(\mathbf{\Lambda}_J(v)\mathcal{U}_J(v))\,,\nonumber
\eeqa
where
\beqa\label{wbar}
\bar w_J(v){\equiv}\sum_{\ell=0}^{L}\,\pint\limits_{4}^{\mu^2}ds\, w_{\ell}(s) k_{\ell,J}(v,s)\,.
\eeqa

Before moving to the dual problem, we observe that using the symmetry $z\to-z$ of the even spin partial-waves $f_{2n}(s)$, we can extend the integration domain of the even spins $w_{2n}$ in the definition of \eqref{wbar} up to $\mu^2_e \leq 60$. This is achieved by integrating over half of the angles $z\in [0,1]$ in \eqref{royequations} and compensating by an overall factor of 2. The region of integration for the odd spins is kept up to $\mu^2_o\le 32$. In the next section and appendix \ref{analytic_constraints} we will find that having $w_{2n}(s>12)\ne0$ and $w_{2n+1}(s>8)\ne0$ is not feasible, so in practice one is forced to take $\mu^2_e \leq 12$ and $\mu^2_o \leq 8$. Yet, the $z\to-z$ symmetry is what allows the even spins bound to be larger than the odd one.

As in the toy problem \eqref{toyproblem}, we define the dual functional by maximizing (minimizing) $\mathcal{L}^+$ ($\mathcal{L}^-$) over the primal variables. 
The Lagrangian is linear in the primal variables and the extremization tedious, but straightforward -- see appendix \ref{appendix_dualSDP} for the details. The final result is
\beq
\!\!\!\!\!\!\! \min_{\scriptsize\bea{c}\mathbf{\Lambda}^\pm_J \succeq 0\\ \pm \bar w_J(s) \geq 0 \eea}\!\!\!\!\!\!\! \;\;\pm D^{\pm}\equiv \!\!\!\!\!\!\! \min_{\scriptsize\bea{c}\mathbf{\Lambda}^\pm_J \succeq 0\\ \pm \bar w_J(s) \geq 0 \eea}\!\!\!\!\!\!\! \;\;\sum_{n=0}^{{\left \lfloor{L/2}\right \rfloor}}\int\limits_4^{\mu^2_e}\!ds\, X_{2n}(s)\,,
\label{dualproblem}
\eeq
where $J$ is an even spin. Here, $\mathbf\Lambda_J$ is the dual matrix that is associated with the unitarity constraint. It is given by
\beq
\mathbf{\Lambda}_J^\pm=\begin{pmatrix}
X_J & \mp \frac{w_J}{2\rho}\\
\mp \frac{w_J}{2\rho} & \frac{1}{4}(\rho^2 X_J \pm 2 \bar w_J)
\end{pmatrix}\succeq 0\,.
\label{dualunitarity}
\eeq
In (\ref{dualproblem}) this constraint is imposed for $4<s<\mu^2_e$ and $0\leq J\leq L$. In the complementary set, the following linear constraints are imposed
\beq\label{dualpositivity}
\pm \bar w_{J\le L}(s>\mu^2_e)\geq 0\ ,\qquad \pm \bar w_{J> L}(s>4)\geq 0\ .
\eeq

Finally, 
the dual variable $w_0$ shall be normalized to
\beq\la{normalization}
\int\limits_4^{\mu^2_e}ds\, w_0(s)=\frac{1}{2}\,.
\eeq

\subsection{Numerical results}
\label{numerical_results}

\begin{figure}[t]
\centering
        \includegraphics[scale=0.283 ]{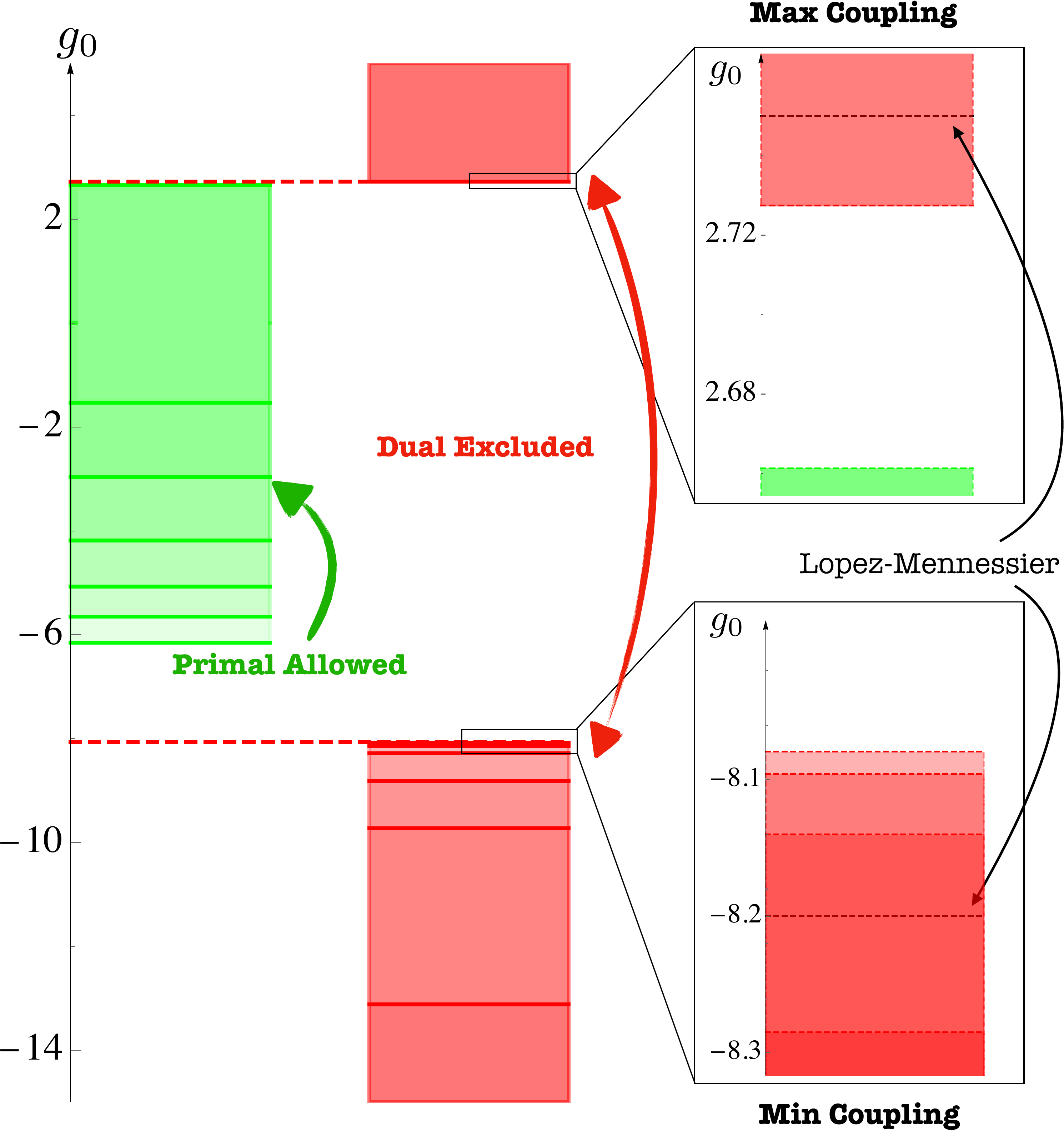}
\caption{Bounds on the quartic coupling $g_0$. On the left, we show in green the region determined by solving the primal problem, in red the one rigorously excluded by solving the dual \eqref{dualproblem}. 
The bound on the maximum coupling converges fast both in the primal and dual problem and the gap is relatively small, though non-vanishing -- right-top inset. 
The bound on the minimum coupling is hard to study using primal: the different green lines denote the numerical coupling for $N_\text{max}=5,8,11,14,17,20$.
On the contrary, the red lines for $L=0,{\dots},6$ show that dual convergence is achieved faster -- right-bottom inset.
In dashed black we report the best values of \cite{Lopez:1976zs}.} 
\label{summary}
\end{figure}

We have implemented the dual problem (\ref{dualproblem}\,-\,\ref{normalization}) in $d=4$ numerically. The summary of this investigation together with the one for the primal problem is plotted in figure \ref{summary}. 
In red we depict the rigorously excluded region for $g_0$. In green, the allowed region obtained using primal numerics is indicated for comparison.

\paragraph*{Primal -}We solve the primal problem by considering a manifestly crossing symmetric ansatz of the form
\beq
T^\text{primal}(s,t,u)=\frac{c}{\rho_s-1}+\!\sum_{a+b\leq N_\text{max}}\!\!\!c_{(ab)}\rho_s^a \rho_t^b +\text{symm}\,,
\label{primalansatz}
\eeq
with $\rho_x=(\sqrt{8/3}-\sqrt{4{-}x})/(\sqrt{8/3}+\sqrt{4{-}x})$ and by imposing unitarity numerically up to some spin $L_\text{max}$ on a grid of points.
For the maximum coupling case, convergence is remarkably fast.
Conversely, the minimum coupling convergence is terrible: the different green lines in figure \ref{summary} correspond to increasing values of $N_\text{max}$ indicating that primal convergence is far from being attained.
\begin{figure}[t]
\centering
        \includegraphics[scale=0.31 ]{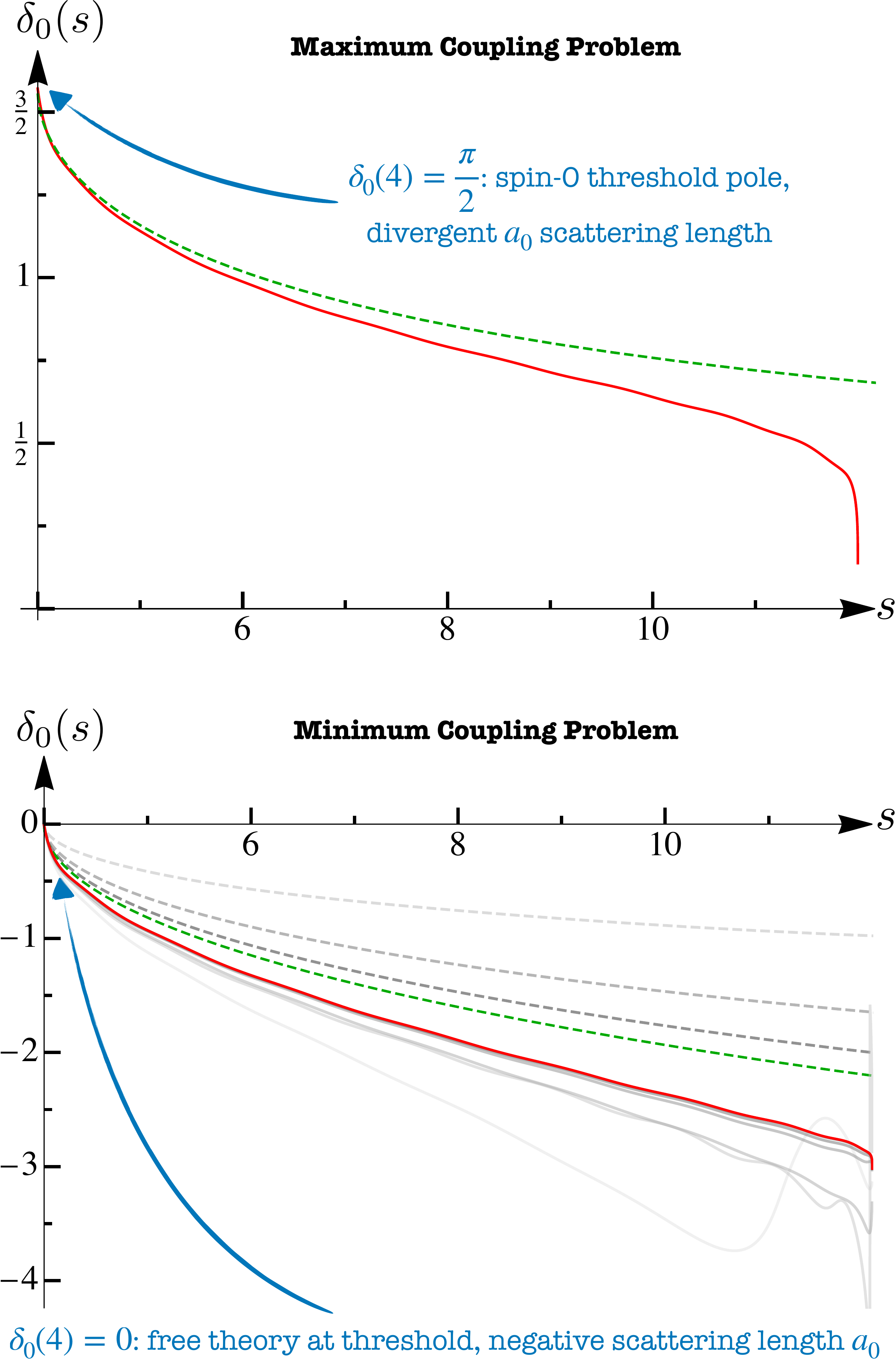}
\caption{
The spin-$0$ phase shifts, $\delta_0=\tfrac{1}{2i}\log S_0$, as a function of the center of mass energy $s$ for the maximum (top) and minimum coupling (bottom). We plot in dashed-green and solid-red the phase shifts obtained respectively from the best primal and dual numerics. In the bottom figure, the dashed lines in gray-scale correspond to increasing values of $N_\text{max}$ up to $N_\text{max}=20$ (in green); the solid lines in gray-scale are obtained from the dual for different values of $L$, up to $L=6$ (in red). Although primal and dual results still differ, the physical content they describe is the same, converging one towards the other as the duality gap shrinks. } 
\label{phaseshifts}
\end{figure}

\paragraph*{Dual-}

The functional constraints (\ref{dualproblem}\,-\,\ref{normalization}) are implemented as follows. The dual variables $X_{2n}$ and $w_\ell$ for $n=0,\dots, {\left \lfloor{L/2}\right \rfloor}$ and $\ell=0,\dots,L$ are parametrized using a simple basis of functions that we took to be the Chebyschev polynomials. We then choose a grid of points where we impose \eqref{dualunitarity} and \eqref{dualpositivity}, {see appendix \ref{dual_numerics_appendix} for details.}

The set of constraints (\ref{dualpositivity}) on the sign of $\bar w_{J>L}$ is however infinite because the spin is unbounded. To implement them numerically, we first trivialize them at large $J$. This leads to a bound on the integration domain $\mu_e^2\leq12$, $\mu_o^2\leq8$ -- see appendix \ref{analytic_constraints} for the details. We then introduce a spin cutoff $J_\text{max}$ on the set (\ref{dualpositivity}) and gradually increase it. At intermediate spins larger than $J_\text{max}$ we have also implemented the sign constraint near the two-particle threshold, see appendix \ref{signthresholdapp}. For the maximal coupling problem, we observe that beyond a certain value there are no dual constraint violations. For the minimal coupling problem, we always have some tiny violations at some $J>J_\text{max}$, hows effect on the bound is negligible.

For the dual maximum coupling the simplicity of the primal mirrors into the dual. An almost optimal bound is attained using just $L=0$, and adding further dual variables $w_\ell$ does not improve significantly the bound. For the minimum coupling, the dual numerical convergence is relatively slower as in the primal case. The red lines in figure \ref{summary} are obtained adding multipliers from $L=0$ up to $L=6$. However, we stress again that for any fixed $L$ the  bounds obtained with large $J_\text{max}$ convergence are rigorous. Increasing further $L$ will possibly make the duality gap smaller. \footnote{See appendix \ref{dual_numerics_appendix} for a detailed analysis of the dual numerics.}

The different convergence rate of the two problems can be understood as follows. The quartic coupling $g_0$ can be measured using the dispersion relation $\mathcal{A}(4,0)$ in \eqref{dispersion} subtracting at $s_0=t_0=4/3$
\beqa
32 \pi g_0=\Re T(4,0)&&-\frac{1}{\pi}\pint\limits_4^\infty dv [T_v(v,0)\,K(v,4,0;4/3)\nonumber\\
&&+T_v(v,4/3)K(v,0,4/3;4/3)]\,.
\label{disp_coupling}
\eeqa
The integrand on the right hand side is positive since $T_v(v,t_0)\geq 0$ for any $t_0 \geq 0$. 
From \eqref{disp_coupling} it is evident that maximizing the coupling is equivalent to minimizing the imaginary part (or the total cross-section).
On the other hand, when we minimize $g_0$ the optimal solution will have a total cross-section that is as big as possible compatibly with unitarity.
In this sense it is not surprising that primal convergence for the minimum coupling case is so hard since the imaginary part of our ansatz \eqref{primalansatz} does not grow at fixed $N_\text{max}$. \footnote{It might be worth investigating whether the amplitude minimizing the quartic coupling also saturates the Froissart bound in $d=4$.} 

In figure \ref{phaseshifts} we compare the primal and dual  spin-$0$ phase shift for the maximum and minimum coupling problem. The phase shifts for both problems tend to differ less as the duality gap shrinks. Moreover, it is encouraging to notice they have the same threshold behavior: the amplitude saturating the maximum coupling has a threshold singularity; the one saturating the minimum coupling a negative scattering length -- see appendix \ref{th_unitarity} for further details.

\section{Discussion and outlook}

In this Letter, we proposed a dual approach to the $S$-matrix Bootstrap based solely on the proven analyticity properties of scattering amplitudes.\footnote{See \cite{Mizera:2021ujs,Mizera:2021fap} for a recent derivation of crossing symmetry using on-shell methods.}
Our strategy consists of decomposing both crossing and unitarity into simpler constraints that can be systematically added to improve the bounds.

Still, there are some important questions to address.
The first one concerns the duality gap. 
The dual problem we optimized numerically \eqref{dualproblem} is not the `mathematical dual' of the primal problem we solved following \cite{Paulos:2018fym} and the duality gap does not necessarily close. 
In particular, in the dual, we imposed full unitarity up to $s=12$.
This limitation follows from the dispersion relations we assumed in (\ref{dispersion}).\footnote{After Roy's papers other authors have tried to further extend the domain of validity of the Roy equations \cite{Mahoux:1974ej, Auberson:1974in,Auberson:1977re}, see \cite{Sinha:2020win,Gopakumar:2021dvg} for recent studies.} 
It might be interesting to use more refined dispersion relations  \cite{work_in_progress} and check whether the gap can be further shrunk. In particular, it would be interesting to extend the applicability of the dual method beyond $s=16$, where non-elasticity is expected to kick in.

Although not proven, maximal analyticity is a typical working assumption made in bootstrap studies. 
It would be worth repeating our analysis under such a hypothesis and compare it with our rigorous bounds. It would also be interesting to extend the dual formulation to the case of massless particles. Clearly, in that case, the validity regime of our formulation shrinks to zero and some assumptions are needed in order to make progress. Related to that, it would be important to generalize the dual problem we formulated to bound Wilson coefficients in EFTs that recently has received a lot of attention \cite{Arkani-Hamed:2020blm,Green:2019tpt,Bellazzini:2020cot,Tolley:2020gtv,Caron-Huot:2020cmc,Caron-Huot:2021rmr,Bern:2021ppb,Kundu:2021qpi}. 

Finally, one of the hardest challenges in the $S$-matrix Bootstrap program concerns the inclusion of multi-particle processes. 
The proliferation of Mandelstam invariants is the bottleneck of manifestly crossing symmetric approaches.
Single variable dispersion relations can overcome this issue. Whether the dual technology developed in this Letter can be used to tackle such challenging problems is an open question we think it will be important to address.

\begin{acknowledgments}
We thank M. Correia, J. Elias-Mir\'o 
, A. Homrich, J. Penedones, A. Raclariu, P. Vieira, and A. Zhiboedov for useful discussion. We thank P. Vieira for comments on the draft. AG was supported by The Israel Science Foundation (grant number 2289/18). AS was supported by the Israel Science Foundation (grant number 1197/20).
\end{acknowledgments}

\appendix

\section{Fixed-$t$ dispersion relations}
\label{disp_derivation}

In this appendix, we give a derivation of the fixed-$t$ dispersion relation with two subtractions in \eqref{dispersion}.
For real $t<4$ the amplitude is polynomially bounded by $(s/s_0)^2$, \cite{Martin2}. Hence, there are only two subtractions needed for a dispersion representation. It is useful to perform these subtractions using a simple division by $s^2$. In that way, we arrive at
\beqa
T(s,t)&=&{s^2\over 2\pi i}\oint{dv\over v-s}{T(v,t)\over v^2}\la{dispersion0}\\
&=&T(0,t)+s\,T'(0,t)+{s^2\over\pi}\int\limits_{4}^\infty{dv\over v-s}{T_s(v,t)\over v^2}\nn\\
&+&{s^2\over\pi}\int\limits_{-\infty}^{-t}{dv\over v-s}{T_v(v,t)\over v^2}\nn\\
&=&T(0,t)+s\,T'(0,t)+{s^2\over\pi}\int\limits_{4}^\infty{dv\over v-s}{T_v(v,t)\over v^2}\nn\\
&+&{s^2\over\pi}\int\limits_{4}^\infty{dv\over v-u}{T_v(v,t)\over (4-t-v)^2}\nn
\eeqa
where $T'(0,t){=}\partial_v T(v,t) |_{v=0}$ and $T_s(s,t){=}{\rm Disc}_sT(s,t)$, and in the second step we have used crossing symmetry.
This $t$-channel dispersion integral representation is however not manifestly $s\leftrightarrow u$ crossing symmetric. To make this symmetry manifest, we express the integral as 
\beqa
&&T(s,t)=T(0,t)+s\,T'(0,t)+2{s^2\over\pi}\int\limits_{4}^\infty{dv\over v-s}{T_v(v,t)\over v^2}\nn\\
&&+{s^2\over\pi}\int\limits_{4}^\infty dv\,T_v(v,t)\[{1\over(v{-}u) (4-t-v)^2}-{1\over (v{-}s)v^2}\]\,.\nn
\eeqa
After averaging this with the $u$-channel we arrive at \cite{Roy:1971tc}
\beq
T(s,t)=g(t)+{1\over\pi}\int\limits_{4}^\infty \!dv\,{T_v(v,t)\over v^2}\[{s^2\over v-s}+{u^2\over v-u}\]\ ,
\eeq
where
\beq
g(t)=T(0,t)+{4{-}t\over2}\,T'(0,t)+{1\over2\pi}\int\limits_{4}^\infty\! dv\,{T_v(v,t)\over v^2}{(4{-}t)^3\over (4{-}t{-}v)^2}\ .
\eeq

Next, we want to trade $g(t)$ for a known function of $t$ and an unknown subtraction constant.
We use first the crossing equation $T(t,t_0)=T(t_0,t)$ to eliminate $g(t)$ in terms of $g(t_0)$ with $0<t_0<4$
\beqa
g(t)=&&g(t_0){-}\frac{1}{\pi}\int\limits_{4}^\infty\! dv\,\frac{T_v(v,t)}{v^2}\left( \frac{t_0^2}{v{-}t_0}+\frac{(4{-}t{-}t_0)^2}{v-4+t_0+t} \right)\nonumber\\
&&{-}\frac{1}{\pi}\int\limits_{4}^\infty \!dv\,\frac{T_v(v,t_0)}{v^2}\left( \frac{t^2}{v{-}t}+\frac{(4{-}t{-}t_0)^2}{v{-}4{+}t_0{+}t} \right)\,,
\eeqa 
and then using the definition of $T(s_0,t_0)$ with $0<s_0<4$ (so that $T(s_0,t_0)$ is real) we eliminate also $g(t_0)$, obtaining the equation in \eqref{dispersion}.

Note that a function that satisfies this $\mathcal{A}(s,t)=0$ constraint is manifestly $s\leftrightarrow u$ crossing symmetric, but not necessarily $s\leftrightarrow t$ symmetric. 

\section{Derivation of the SDP problem}
\label{appendix_dualSDP}

In this appendix we derive the dual problem (\ref{dualproblem})-(\ref{normalization}) by extremizing the Lagrangian (\ref{finalLagrangian}) over the primal variables, $f_\ell(s)$, and some of the components of the dual variables $\mathbf{\Lambda}_J(s)$.

The integrand of the unitarity term can be expanded as
\beqa
\text{tr}(\mathbf{\Lambda}_J(s) \mathcal{U}_J(s))&{=}&2 \Im f_J(s) \mathbf{\Lambda}^{22}_J(s){+}2 \Re f_J(s) \rho^2(s)\mathbf{\Lambda}^{12}_J(s)\nn\\
&+&\mathbf{\Lambda}^{11}_J(s)\left(1-\rho^2(s)\Im f_J(s)/2\right)\,.
\label{unitarity_term}
\eeqa

Extremizing (\ref{finalLagrangian}) w.r.t. the primal variables yields the equations 
\beqa
(\mathbf{\Lambda}_J^{22})^\pm&=&\left[ \rho^2(s)\mathbf{\Lambda}_J^{11}(s) \pm 2\bar w_{ J}(s) \right]/4\,,\\
(\mathbf{\Lambda}_J^{12})^\pm&=&\left\{\begin {array}{lc}\mp \frac{w_J(s)}{2\rho(s)}&\quad (4<s<\mu^2)\,\land\,(\text{even }J \leq L)\\
0&\text{otherwise}\end{array}\right..\nn
\label{solinner}
\eeqa
We use these equations to eliminate $\mathbf{\Lambda}_J^{12}$ and $\mathbf{\Lambda}_J^{22}$. After this choice, the Lagrangian takes the form 
\beq\label{dual_objective_appendix}
D^\pm(s_0,t_0)\equiv \pm \sum_{J\,\text{even}}\int\limits_4^{\mu^2_e}\!ds\, X_{2n}(s)\ ,
\eeq
where $X_J(s)\equiv \mathbf{\Lambda}_J^{11}$. It is subject to the semi-definite positive condition 
\beq\label{dualunitmatrices}
\mathbf{\Lambda}^\pm_J(s)=\begin{pmatrix}
X_J(s) & \mp \frac{w_J(s)}{2\rho(s)} \\
\mp \frac{w_J(s)}{2\rho(s)} & \frac{1}{4}\left( \rho^2(s)X_J(s)\pm 2 \bar w_{J}(s) \right)
\end{pmatrix}{\succeq} 0\ ,
\eeq
for $(4<s<\mu^2_e)\land(J\leq L)$ and 
\beq\label{thetasdp}
\mathbf{\Lambda}^\pm_J(s)=\begin{pmatrix}
X_J(s) & 0 \\
0 & \frac{1}{4}\left(\rho^2(s)X_J(s) \pm 2  \bar w_{\ell}(s) \right)
\end{pmatrix}{\succeq} 0\,,
\eeq
otherwise. The equation of motion for $g_0$ results in the additional normalization condition (\ref{normalization}).

At this point, the dual objective is still an infinite sum of dual variables, making the problem difficult to solve. However, the constraint in (\ref{thetasdp}) can be solved analytically by setting $X_J(s)=0$ its domain of validity $\((s>\mu^2_e)\land(J\leq L)\) \lor\((s>4)\land(J>L)\)$, 
and imposing the simpler linear conditions
\beq\label{linearfinal}
\pm   \bar w_{J}(s) \geq 0\,,
\eeq
in this regime. This last step leads to the final formulation of the \emph{Dual SDP Problem} quoted in the main text (\ref{dualproblem}).

\section{Derivation of the non-linear problem}
In this appendix, we derive the dual problem by imposing unitarity in the form of the non-linear inequality $\det \mathcal{U}_\ell(s) \geq 0$.
This form has some conceptual advantages, but it is difficult to solve numerically. 
We will not use it for any systematic numerical exploration, but it will give us a procedure to extract the primal phase shifts from the dual SDP problem \eqref{dualproblem}.

We start from the Lagrangian in (\ref{finalLagrangian}) and replace the SDP unitarity constraint by its non-linear form
\beq
\tr(\mathbf{\Lambda}_J(s)\mathcal{U}_J(s))\ \rightarrow\  \lambda_J(v) (2\Im f_J(s)-\rho^2 |f_J(s)|^2)\ .\nn
\eeq

The equation of motions for the primal variables $\Re f_J$ and $\Im f_J$ yield
\beqa\label{eomNL}
\Im f_{J}(s)&=&\frac{1}{\rho^2(s)}\mp \frac{\bar w_{J}(s)}{2 \rho^2(s)\lambda_{J}(s)}\,,\\
\Re f_J(s)&=&\pm\left\{\!\begin {array}{lc}\frac{w_J(s)}{2 \rho^2(s)\lambda_J(s)}& (4<s<\mu^2_e)\land(\text{even }J \leq L)\\
0&\text{otherwies}\end{array}\right.\ \nn  
\eeqa

As before, the equation of motion for $g_0$ implies (\ref{normalization}).

By plugging the solution \eqref{eomNL} into the Lagrangian we arrive at the dual function
\beqa
\mathcal{D}^\pm(s_0,t_0)\equiv& \pm&\sum_{n=0}^{{\left \lfloor{L/2}\right \rfloor}} \int\limits_{4}^{\mu^2_e}dv\, \frac{w^2_{2n}(v)}{4 \rho^2(v)\lambda_{2n}(s)} \\
&\pm& \sum_{J\,\text{even}}^\infty \int\limits_{4}^\infty dv\, \frac{(\bar w_{J}(v) \mp 2 \lambda_J(v))^2}{4\rho^2(s)\lambda_J(s)}\nn\,.
\eeqa

The dual unitarity variables $\lambda_J(s)$ are unconstrained positive functions and they appear quadratically in the dual objective.
It is, therefore, possible to find the extremum of this action analytically w.r.t. $\lambda_J$.

For $4<s<\mu^2$ and $J\leq L$ we get
\beq
\lambda_J(s)=\frac{1}{2}\sqrt{w^2_J(s)+ \bar w_{J}^2(s)}\geq 0\,,
\label{lambdainner}
\eeq
and in the complementary region
\beq
\lambda_J(s)=\mp \frac{1}{2} \bar w_{J}(s)\,, \qquad \lambda_J(s)\geq 0\,.
\label{lambdaouter}
\eeq

Plugging the solution \eqref{lambdainner} and \eqref{lambdaouter} into the dual action yields
\beqa
\label{dualNLfinal}
\tilde{\mathcal{D}}^\pm(s_0,t_0)&\equiv&
\sum_{n=0}^{\left \lfloor{L/2}\right \rfloor} \int\limits_{4}^{\mu^2_e}\!ds\, \frac{\left( {-}\bar w_{2n}(s)  \pm \sqrt{w^2_{2n}(s)+ \bar w^2_{2n}(s)} \right)}{\rho^2(s)}\nn\\
&-&\sum_{n=0}^{\left \lfloor{L/2}\right \rfloor}\int\limits_{\mu^2_e}^\infty\!ds\, \frac{2}{\rho^2(s)} \bar w_{2n}(s) \theta\left( \mp \bar w_{2n}(s) \right)\\
&-&\sum_{n>{\left \lfloor{L/2}\right \rfloor}}\int\limits_{4}^\infty\!ds\, \frac{2}{\rho^2(s)} \bar w_{2n}(s) \theta\left( \mp \bar w_{2n}(s) \right)\,.\nn
\eeqa

The dual problem in terms of $\tilde{\mathcal{D}}^\pm$ is nonlinear and unconstrained modulo the spin zero  normalization condition (\ref{normalization}). However, the presence of the step function makes this non-linear formulation challenging to solve using gradient methods.\footnote{It is possible to replace an integrand of the form $-x\,\theta(-x)$ with its smooth approximation using the inequality $-x\,\theta(-x)=-x+\sqrt{x^2}\leq -x+\sqrt{x^2+\epsilon^2}$ for some $\epsilon$ in order to use simple gradient algorithms. It might be worth exploring this possibility. A similar trick works also for $-x\,\theta(x)$.}

The dual objective \eqref{dualNLfinal} can be further simplified by imposing the inequality $\pm \bar w_{J}(s) \geq 0$ on the support of the $\theta$ functions. Then the $\theta$ functions contributions can be neglected and this is consistent respectively with the minimization/maximization of $\tilde{\mathcal{D}}^\pm(s_0,t_0)$. 
These conditions coincide with those we find in the SDP case \eqref{linearfinal}. The non-linear dual problem for the quartic coupling becomes
\beqa
\label{dualNLfinal2}
&&\,\min  \; \pm \tilde{\mathcal{D}}^\pm(s_0,t_0)\\
=&&\,\min  \; \sum_{n=0}^{\left \lfloor{L/2}\right \rfloor} \int\limits_{4}^{\mu^2_e}\!ds\, \frac{ \mp \bar w_{2n}(s)  + \sqrt{w^2_{2n}(s)+ \bar w^2_{2n}(s)} }{\rho^2(s)}\,,\nn
\eeqa
subject to the constraints \eqref{linearfinal}.

The problem \eqref{dualNLfinal2} depends on a finite number of dual functions $w_\ell(s)$ for any $0\leq \ell\leq L$, but it is non-linear and constrained.
It would be important to develop an efficient numerical algorithm to solve this class of problems.

\section{Phase shifts}\la{phapp}

In the SDP formulation of the Bootstrap problem \eqref{dualproblem}, we cannot reconstruct the phase shifts once we extremize over the primal variables.
On the other hand, in the non-linear formulation \eqref{dualNLfinal}, using the eqs. \eqref{eomNL} and \eqref{lambdainner} it is straightforward to obtain 
\beq
S_J^\pm(s)=1+i \rho^2(s)f^\pm_J(s)=\mp \frac{ W^*_J}{|W_J|}\,,\quad W_J=i w_J-\bar w_J,
\label{dualphaseshifts}
\eeq
respectively for the maximum and minimum coupling problem and for $(4<s<\mu^2_e)\land(0 \leq J \leq L)$.

It does not come as a surprise that the two problems are simply related since they were derived using different equivalent versions of the unitarity constraints.
To understand their relation, we look at the dual unitarity constraints \eqref{dualunitmatrices}. Suppose $\det \mathbf{\Lambda}^\pm_J(s)=\epsilon^2>0$, then we can solve for $X_J>0$ obtaining
\beqa
X_J^*(s)&=&\frac{\pm \bar w_J+\sqrt{\bar w^2_J +w_J^2+\epsilon^2}}{\rho^2}\geq \frac{\pm \bar w_J+\sqrt{\bar w^2_J+w_J^2}}{\rho^2}\nn\\ \la{D2}
\eeqa
The last inequality implies an inequality for the linear and non-linear dual objectives
\beq
|D^\pm(s_0,t_0)|\geq |\tilde{\mathcal{D}}^\pm(s_0,t_0)|\,,
\eeq
with the equality attained when $\epsilon=0$, or equivalently, when $\det \mathbf{\Lambda}^\pm_J(s)=0$.

In that case, when $\epsilon=0$, the two objectives coincide and we can use the definition \eqref{dualphaseshifts} and apply it to the numerical solution of our SDP problem.
In practice, $\det \mathbf{\Lambda}^\pm_\ell(s)\approx 0$ for all our numerical results -- see figure \ref{dualps_max} and figure \ref{dualps_min} for a comparison between the reconstructed phase shifts using (\ref{dualphaseshifts}) and the direct primal phase shifts.

\section{Analytic constraints from asymptotic regimes}
\label{analytic_constraints}

The sign constraints on the $\bar w_{J}(v)$ functionals (\ref{dualpositivity}) should be imposed to an arbitrary high spin. This is, of course, outside the scope of a numerical application. To overcome this difficulty, in this appendix we solve these constraints analytically at large spin. Doing so is crucial for our numerical implementation to converge into a solution that satisfies all the constraints at large $J_\text{max}$. We find that the large $J$ constraints are only feasible if  $\mu^2_e\le12$ and $\mu_o\le8$. For simplicity, in this appendix, we set $d=4$.

Recall the definition of $\bar w_{J}(v)$ in (\ref{wbar}) that we repeat here for convenience
\beq
\bar w_J{\equiv}\sum_{n=0}^{{\left \lfloor{L/2}\right \rfloor}}\; \pint\limits_{4}^{\mu^2_e}ds\, w_{2n} k_{2n,J} {+}\sum_{n=0}^{\left \lceil{L/2{-}1}\right \rceil}\; \pint\limits_{4}^{\mu^2_o}ds\, \frac{w_{2n{+}1}}{2} k_{2n{+}1,J}\,,
\label{wbardetailed}
\eeq
where, $\mu^2_e\leq 60$ and $\mu^2_o\leq 32$. The dependence on $J$ of $\bar w_J(v)$ only comes from the kernels
\beqa\label{kellJ}
&&k_{\ell,J}(v,s)=\frac{2J+1}{\pi}\int\limits_{z_\text{min}}^1\! dz\,  P_\ell(z) \times\\
&&[P_{J}(x(v,t))K(v,s,t;t_0)+ P_{J}(x(v,t_0))K(v,t,t_0;s_0)]\,, \nonumber
\eeqa
where $z_\text{min}=0$ for $\ell$ even, $z_\text{min}=-1$ when $\ell$ is odd,  $t=t(z)=\tfrac{1}{2}(s-4)(z-1)$, and $P_\ell(z)=P_\ell^{(4)}(z)$ are the Legendre polynomials. To solve the constraint on $\bar w_J$ at large $J$ we will simplify (\ref{kellJ}) into a single $J$-independent constraint. The range of $v$ in which this constraint is imposed can be divided into two regions, inner and outer. In each of these regions one of the two terms in (\ref{kellJ}) dominates over the other.

Consider the second term first. The constraint (\ref{dualpositivity}) is imposed for $v>4$ and $t_0=3/4>0$. In this regime, the argument of the partial wave is larger than one, $x(v,t_0)=1+\frac{2 t_0}{v-4}>1$. At large $J$ and for $x>1$, the Legendre polynomials $P_J(x)$ grow exponentially as 
\beq
P_{J}(x)= \frac{(x+\sqrt{x^2-1})^J}{\sqrt{2\pi J}\sqrt{1+x(\sqrt{x^2-1}-x)}}\(1+\cO(1/J)\)\,.\label{largeJ}
\eeq
Hence, at large $J$ and for a generic point where the kernel $K$ does not have a zero or a pole, the second term in (\ref{kellJ}) dominate over the first, provided that $x(v,t_0)\ge|x(v,t)|$. Because $t(z)<0$, this can fail only when 
\beq
x(v,t(z))=1+\frac{2t(z)}{v-4}=\frac{v-s+z(s-4)}{v-4}\le-1\ .
\eeq
The maximal value of $|x(v,t(z))|$ is archived at $z=z_\text{min}$ and $s=\mu^2$. At the critical point $v_c$ we have $x(v_c,t(\left.z_\text{min})\right|_{s=\mu^2})=-x(v_c,t_0)$ and hence
\beq
 v_c=2{+}\frac{\mu^2}{2}{-}t_0{-}\frac{z_\text{min}}{2}(\mu^2{-}4)\,.
\eeq
In general, $v_c$ will be different for even and odd spins because of the different values of $z_\text{min}$ and $\mu^2$. However, if we choose the upper limit of integration for odd spins in (\ref{wbardetailed}) to be $\mu^2_o=2+\tfrac{\mu^2_e}{2}$, then $v_c$ is the same and is given by
\beq\la{vcritical}
 v_c=2{+}\frac{\mu^2_e}{2}{-}t_0\,.
\eeq
As we now explain, making this choice will allow us to factor out the common $J$-dependant factor from the two terms in (\ref{wbardetailed}). We denote the case where $v>v_c$, \emph{outer region}, in which the term with $P_J(x(v,t_0))$ dominates the integral over $z$ and $s$. In the \emph{inner region}, $4<v<v_c$, and the term with $P_J(x(v,t))$ dominates the integrals. 

The above division into inner and outer regions of dominance can break down at points where one of the kernels in (\ref{kellJ}) has a zero or a pole. However, because the partial waves are exponentially large, the contribution of such isolated points to the integral in (\ref{kellJ}) is negligible. 

\subsection{Constraints from the outer region}
In the outer region $v>v_c$ and at large $J$ equation \eqref{kellJ} becomes
\beq
k_{\ell,J}(v,s)\sim\frac{2J+1}{\pi}P_J\left(x(v,t_0)\right) k_\ell^\text{out}(s,v)\,,
\eeq
with
\beq
k_\ell^\text{out}(s,v)=\int\limits_{z_\text{min}}^1\!dz\,  P_\ell(z)\,  K(v,t(z),t_0;s_0)\,.
\eeq
Using the definition \eqref{wbardetailed}, the functional $\bar w_J$ in this region behaves as
\beq
\bar w_{J}(v)\sim \frac{2J+1}{\pi}P_J\left(x(v,t_0)\right)\bar y(v)\,,
\label{outerfunc}
\eeq
with
\beq
\bar y(v)\equiv\sum_{n=0}^{{\left \lfloor{L/2}\right \rfloor}}\int\limits_{4}^{\mu_e^2} ds\,  w_{2n} k_{2n}^\text{out}{+}\sum_{n=0}^{\left \lceil{L/2{-}1}\right \rceil}\int\limits_4^{\mu^2_o}ds\, \frac{w_{2n{+}1}}{2}k^\text{out}_{2n{+}1}\,. \nonumber
\eeq
Because $x(v,t_0)>1$, the sign constraint (\ref{dualpositivity}) reduces to the $J$-independent linear constraint $\pm \bar y(v) \geq 0$. We impose it for $v>v_c$ choosing a grid in $v$.

\subsection{Constraints from the inner region}
\label{inner_largeJ}
In the inner region $4<v<v_c$ the large $J$ behaviour of $\bar w_J$ is more subtle because $x(v,t(z))$ dependence on $z$, and so it does not simply factor out of the $z$ integral in (\ref{kellJ}). Nonetheless, the partial wave inside the integrand grows exponentially, and therefore the integral in \eqref{kellJ} is dominated by the region where $z\sim z_\text{min}$ and therefore
\beq
\begin{aligned}
k_{\ell J}(v,s)\sim&\frac{2}{\pi}{\frac{s (z_\text{min}-1)+v-4 z_\text{min}}{s-4}}\\
&\times P_J\left(x(v,t_\text{min})\right) k_\ell^\text{inn}(s,v)\,,
\end{aligned}
\eeq
with
\beq
k_\ell^\text{inn}(s,v)\equiv P_\ell(z_\text{min})K(v,s,t_\text{min};t_0)\,,  
\eeq
and $t_\text{min}=t(z_\text{min})=-\tfrac{1}{2}(s-4)(1-z_\text{min})$. 
Using \eqref{wbardetailed} and again the fact that the partial wave in the integrand grows exponentially with $J$, we obtain\footnote{Note that $t_\text{min}$ depends on $s$.}
\beq
\bar w_{J}(v)\sim{2(\mu^2_e-v)^3\over\pi J^2 (\mu^2_e-4)^2 }P_J\big(x(v,2-\tfrac{\mu^2_e}{2})\big)\,\Phi(v)\,,\nn
\eeq
where
\beqa
&&\Phi(v)\equiv K(v,\mu^2_e,2-\tfrac{\mu^2_e}{2};t_0)\times\\
&&\Big[\sum_{n=0}^{{\left \lfloor{L/2}\right \rfloor}} w_{2n}(\mu^2_e) P_{2n}(0)
-\!\!\sum_{n=0}^{\left \lceil{L/2{-}1}\right \rceil}\!\!\tfrac{1}{2}
w_{2n+1}(2+\tfrac{\mu^2_e}{2})P_{2n+1}(-1)\Big]\,.\nn
\eeqa
The sign constraint on $\bar w_J$ now becomes a sign constraint on the $J$-independent function $\Phi(v)$. It dependence on $v$ only through the kernel, while the dual parameters only enter at the points $w_{2n}(\mu_e^2)$ and $w_{2n+1}(2+{\mu_e^2\over2})$. Hence, the sign constraint can only be satisfied over a region in which the kernel  
\beqa
&&K(v,\mu^2_e,2-\tfrac{\mu^2_e}{2};t_0)=\\
&&\frac{(\mu^2_e-t_0)(\mu^2_e-4+2 t_0)}{(v-t_0)(\mu^2_e-v)(2v+\mu^2_e-4)}\times\frac{4v-4-\mu^2_e}{\mu^2_e+4-2v-2t_0}\,,\nonumber
\eeqa
has a fixed sign. The first factor is clearly positive. We then demand that the second factor has a fixed sign in the range $v\in[4,v_c]$. At $v=v_c$ this factor reduces to the manifestly positive value ${3\over8}(\mu_e^2+4)$. At $v=4$ it takes the form $(12-\mu_e^2)/(\mu_e^2-4/3)$. Hence, for the sign constraint (\ref{linearfinal}) to be feasible at large spin, we are forced to take $\mu_e^2\le 12$ and $\mu_o^2\le8$ correspondingly. This constraint on the regime of the dual problem comes from the kernel and is independent of the choice $t_0$. Moreover, it can be shown to hold in any dimension $d\ge3$.  

To summarize, provided that $\mu_e^2\le12$, the large $J$ sign constrain in the inner region reduce to a corresponding $J$ and $v$ independent sign constraint on the sum
\beq
\sum_{n=0}^{{\left \lfloor{L/2}\right \rfloor}} {(-1)^n\Gamma(n+\tfrac{1}{2})\over\Gamma(n+1)}w_{2n}(\mu^2_e)+\!\!\sum_{n=0}^{\left \lceil{L/2{-}1}\right \rceil}\!\!\tfrac{1}{2} w_{2n+1}(2+\tfrac{\mu^2_e}{2})\,,
\eeq
where we have used the explicit value of the Legendre function.

\section{Constraints from the threshold}\la{signthresholdapp}

For intermediate spins we have also imposed the $\bar w_J(v)$ sign constrain (\ref{dualpositivity}) near the two-particle threshold up to some cutoff higher than $J_\text{max}$. In this limit, the constraint slightly simplifies as we now explain. 

As $v\to4^+$ the argument of the partial waves in (\ref{kernels}) blows up and we can approximate these polynomials by their leading power
\beq\label{largex}
\lim_{x\to\infty}P^{(d)}_{J}(x)= {_2}{F_1}\left(\tfrac{1-J}{2},-\tfrac{J}{2},\tfrac{d-3}{2},1 \right) x^J\[1+\cO(1/x)\]\,.
\eeq
Correspondingly, in this limit we have
\beqa
\bar w_{J}(v) &&\sim \frac{\mathcal{N}_d\,n_J^{(d)}}{\pi (v-4)^J}{_2}{F_1}\left(\tfrac{1-J}{2},-\tfrac{J}{2},\tfrac{d-3}{2},1 \right) \times\\
&&\Big(\sum_{n=0}^{{\left \lfloor{L/2}\right \rfloor}} \int\limits_4^{\mu^2_e}ds\, w_{2n}k_{2n,J}^\text{thr}\,{-}\sum_{n=0}^{\left \lceil{L/2{-}1}\right \rceil}\int\limits_4^{\mu^2_o}\frac{w_{2n+1}}{2}k_{2n+1,J}^\text{thr}\Big)\,,\nn
\eeqa
where
\beqa
&& k_{\ell,J}^{\text{thr}}=\int\limits_{z_\text{min}}^1 dz P^{(d)}_\ell(z)(1{-}z^2)^{\tfrac{d{-}4}{2}}\times\\
&& \times [(4t_0)^J K(4,t(z),t_0;s_0)+(4t(z))^J K(4,s,t(z);t_0)]\,.\nonumber
\eeqa

\section{Dual Numerics}
\label{dual_numerics_appendix}

In this appendix we give more details about the numerical implementation.

We choose the following parametrization for the dual variables
\beqa
X_{J}(s)&=&\frac{\alpha^J \delta_{J,0}}{\sqrt{s-4}}+\sum_{n=0}^{N_X(J)} \beta_n^J\, {\rm Ch}_n\big(1+\tfrac{2(s-\mu_e^2)}{\mu_e^2-4}\big)\,,\nn\\
w_\ell(s)&=&\sum_{n=0}^{N_w(\ell)} w_n^\ell\,{\rm Ch}_n\big(1+\tfrac{2(s-\mu_e^2)}{\mu_e^2-4}\big)\,,\label{dualvars_param}
\eeqa
where $\alpha^0, \beta_n^J, w_n^\ell$ are free parameters with $J$ even and $\ell$ any positive integer and ${\rm Ch}_n$ are the Chebyschev polynomials.
The square root singularity at threshold for $X_0$ is allowed by the general threshold expansion of the dual constraints and by the regularity of the dual objective, see (\ref{D2}). It turns out to help the convergence for the maximum coupling problem.

\begin{figure}[t]
\centering
        \includegraphics[scale=0.24 ]{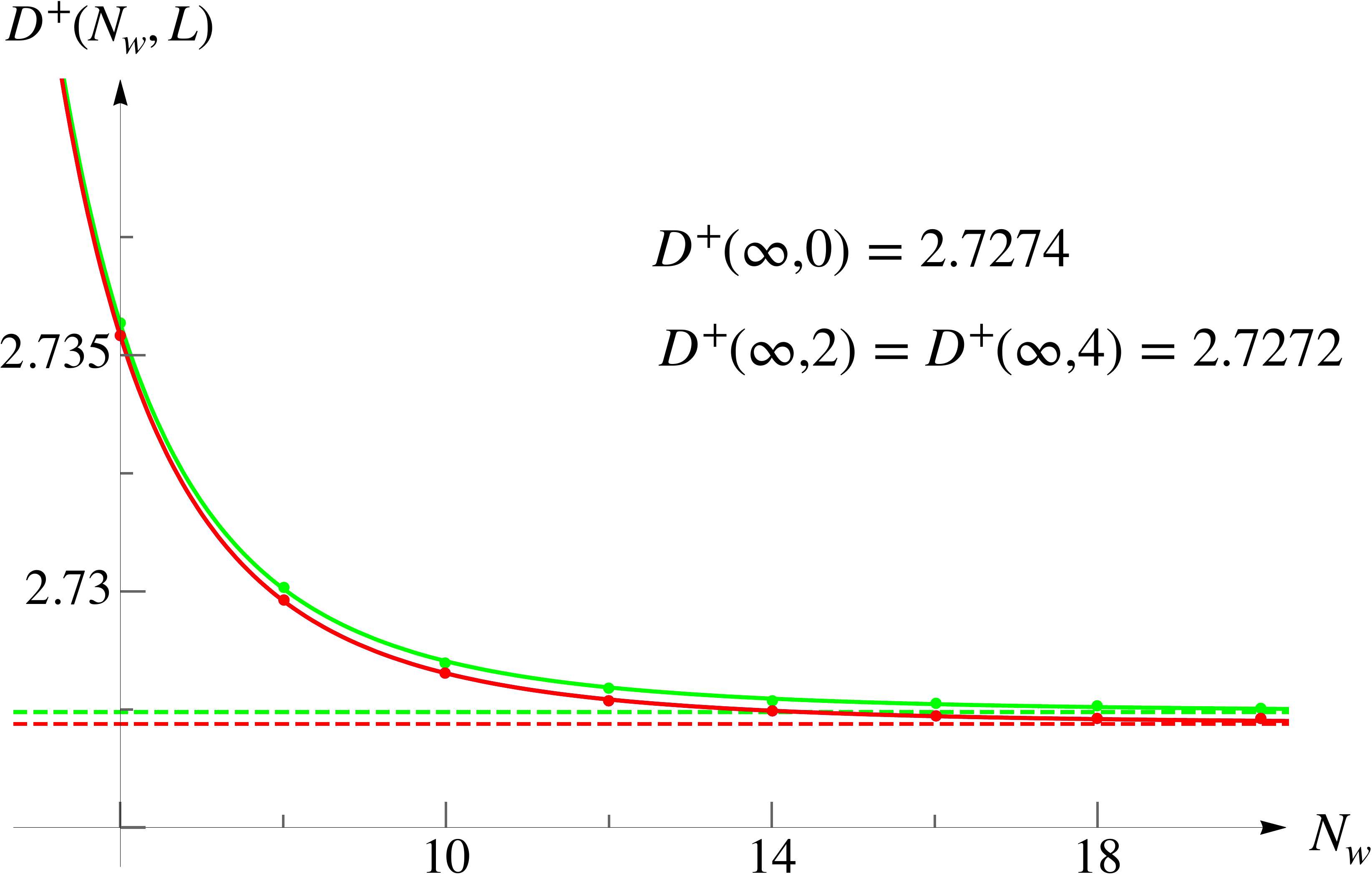}
\caption{Minimum value of the dual objective $D^+(4/3,4/3)$ \eqref{dual_objective_appendix} for the maximum coupling problem as a function of $N_w$ and $L$. Different colors correspond respectively to $L=0$ (green), $L=2$ (blue) and $L=4$ (red). The blue curve is covered by the red one as the bound does improve anymore by increasing $L$. The addition of odd dual variables $w_{2n+1}$ has no effect on the bound. The dashed lines represent the corresponding extrapolations to $L\to \infty$. Our best bound $g_0\leq 2.7272<2.75$ improves the one found in \cite{Lopez:1976zs}, but is still bigger than the one obtained by the primal \cite{Paulos:2018fym} and the dual of \cite{He:2021eqn}.}
\label{max_bound}
\end{figure}

The dual unitarity constraints \eqref{dualunitarity} imply that $X_J(s)\geq 0$. The parametrization in \eqref{dualvars_param} is not manifestly positive and it is convenient to supplement the 
dual constraints with the necessary conditions $X_J(s_i)\geq 0$ for some very refined grid $\{s_i\}$.\footnote{It is possible to trivialize this necessary condition by parametrizing $X_0(s)$ in terms of Bernstein polynomials of the form $(1-x)^k(1-x^j)$. However, we have found such an expansion to converge slowly.}

To run the numerics we choose the following strategy. We fix $N_X(J)$ to some high value, for instance, $N_X(J)=40$  and we slowly increase $N_w(\ell)$.
The reason is that although we need to impose the linear constraints \eqref{linearfinal} for all $J$'s, when running the numerics we need to introduce a cutoff $J_\text{max}$.
The functionals $\bar w_J$ depend only on $N_w(\ell)$, and for any fixed $N_w(\ell)$ we want to make sure that $J_\text{max}$ is large enough to prevent dual constraints violations.
To further simplify the analysis we keep $N_w(2n+1)=4$ fixed. Notice that the $w_{2n+1}$ only enters in the optimization problem integrated against some kernel.
Indeed, $w_{2n+1}$ can wildly oscillate without changing $\bar w_J$. We found empirically that just increasing $N_w(2n+1)$ makes $J_\text{max}$ convergence harder without improving the bound significantly.

The best bound is obtained integrating on the largest possible region in $s$, therefore in our numerics we set $\mu_e^2=12$ and $\mu_o^2=8$ -- see Appendix \ref{inner_largeJ}.
We compute the $\bar w_J$ functionals analytically: the integrated expressions are lengthy but simple as they contain at most logarithms and dilogarithms. 

We map both the region $4<v<\mu^2_e$ and the region $\mu^2_e<v$ to the interval $\[-1,1\]$ and there we impose the dual constraints on a Chebyschev grid with respectively 500 and 80 points. We now discuss our numerical findings in turn for the maximum and minimum coupling problems.

\subsection{Maximum coupling dual problem}

\begin{figure}[t]
\centering
        \includegraphics[scale=0.29 ]{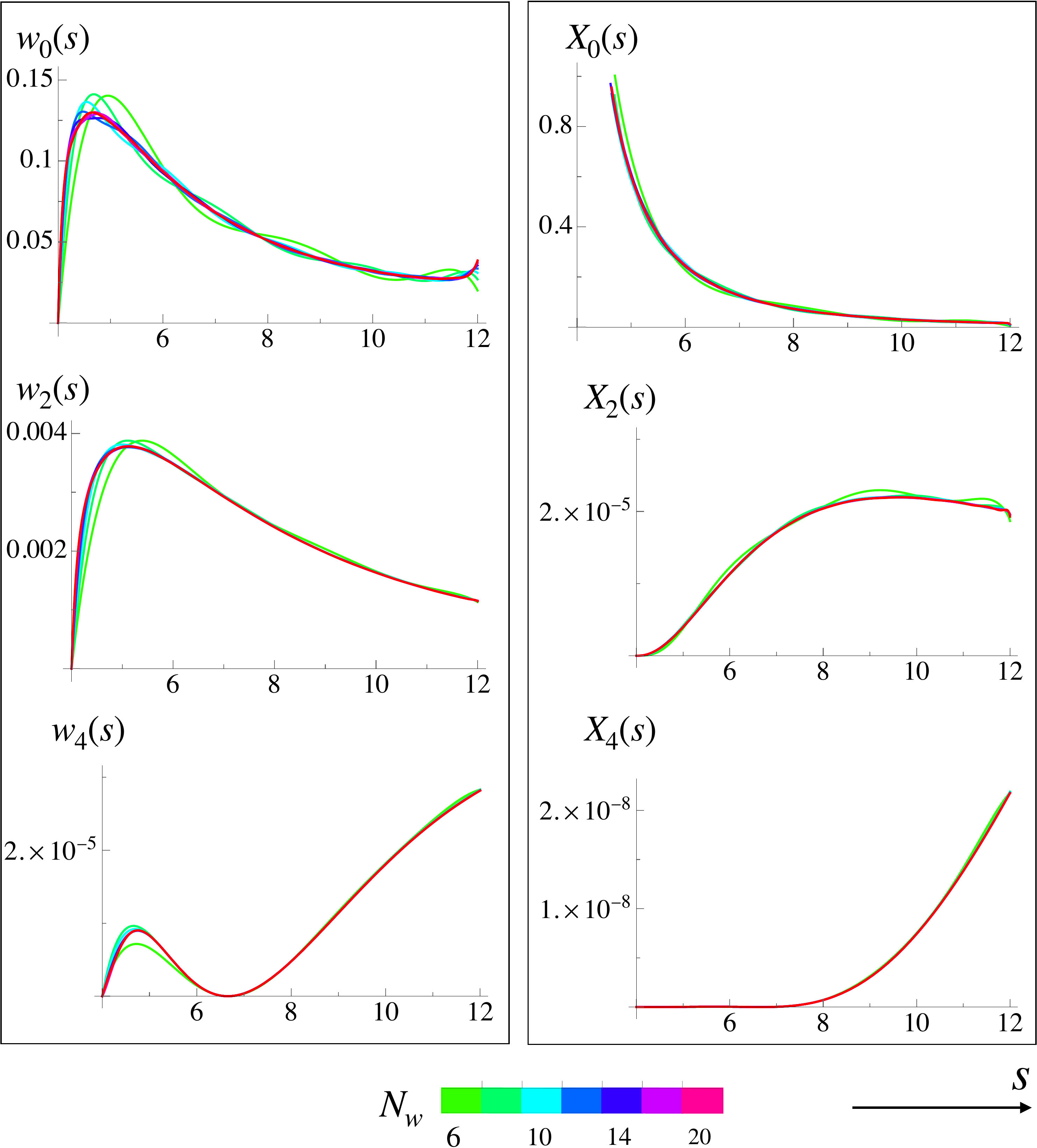}
\caption{In the left panel we show the dual dispersion variables $w_{2n}$ for $n=0,1,2$. On the right, we show the unitarity dual variables. The dual objective $D^+$ is obtained by summing the areas below the $X_J$ curves. We see how small is the contribution to the bound coming from the higher spins. Convergence in $N_w$ is also achieved fast as the various curves tend to overlap by increasing $N_w$.} 
\label{dualvars_max}
\end{figure}

We start with the maximum coupling problem since it turns out to be one of the simplest S-matrix Bootstrap problems. The bound converges already when $L=0$ and adding higher spin constraints does not improve it significantly. We observe that just imposing the $\bar w_2\geq 0$ together with the large $J$ constraints is sufficient to make all the $\bar w_J\geq 0$ constraints with $J>2$ automatically satisfied. Moreover, turning on the odd $w_{2n+1}$'s has an almost zero impact on the bound. 

In figure \ref{max_bound} we plot the value of the minimum value of the dual objective \eqref{dual_objective_appendix} $D^+(4/3,4/3)$, that we simply denote by $D^+$, as a function of $N_w$ for $L=0,2,4$ respectively in green, blue and red (the blue points are invisible since overlap with the red ones).\footnote{We keep $N_w(\ell)$ the same for all even $\ell$.} The dashed lines are the power-law extrapolations for $N_w\to \infty$ with the extrapolated value reported in the figure. Taking the best value we conclude that $g_0 \leq 2.7272$.

\begin{figure}[t]
\centering
        \includegraphics[scale=0.33 ]{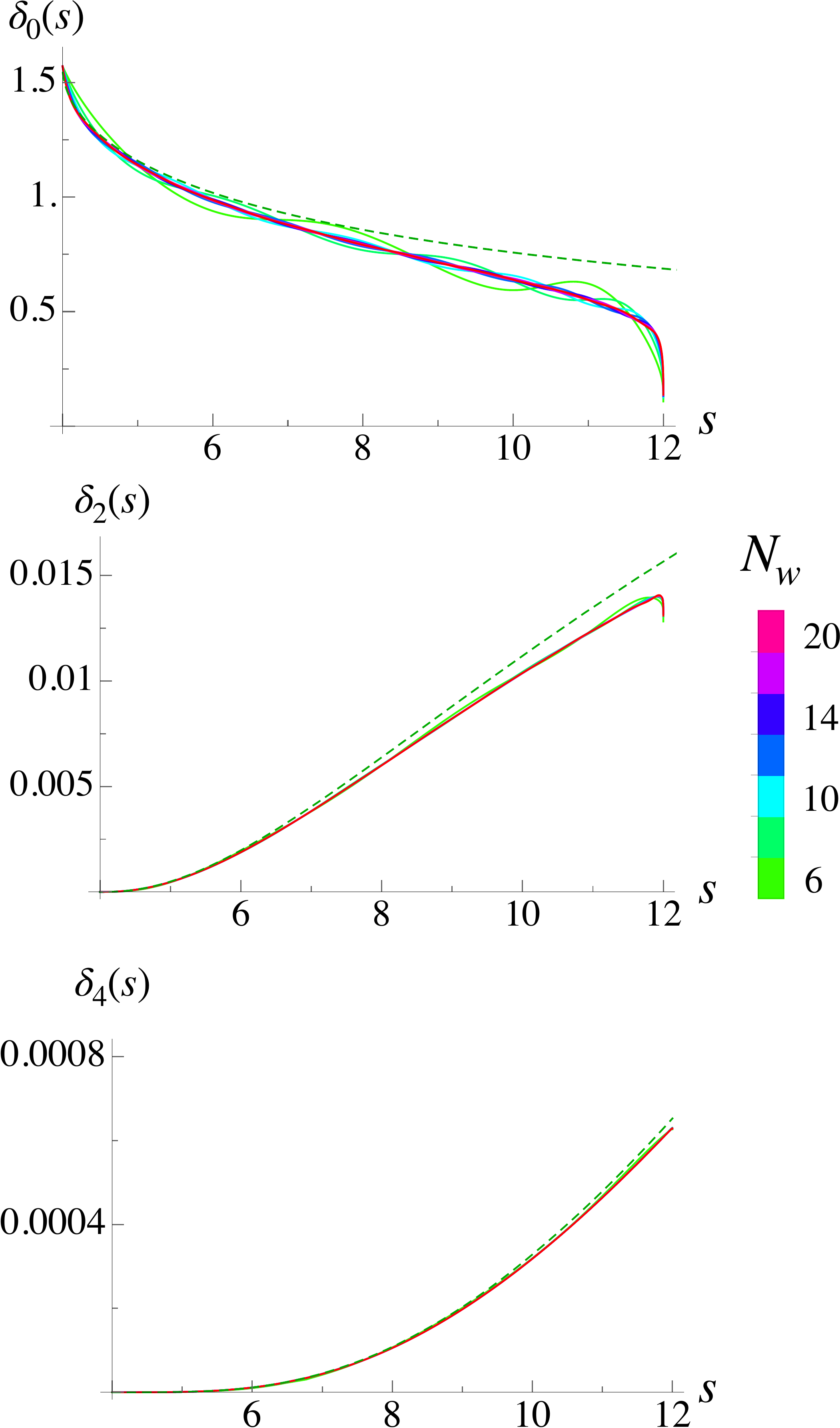}
\caption{Phase shifts $\delta_\ell(s)=\tfrac{1}{2i}\log{S_\ell(s)}$ for the first few spins as a function of $s$. In green dashed we plot the primal phase shifts for our best numerics $N_\text{max}=10$, $L_\text{max}=10$. Solid lines in color gradient are obtained from the dual numerics with $L=4$ and different $N_w$. It is not surprising that primal and dual curves agree as the relative duality gap is relatively small $\Delta_\text{gap}=2\tfrac{|g_0-D^+|}{g_0+D^+}=0.02$.} 
\label{dualps_max}
\end{figure}

In figure \ref{dualvars_max} we show how the various dual variables converge as we increase $N_w$ from $N_w=6$ to $N_w=20$ in color gradient for the $L=4$ numerics. On the left panels we show the dual dispersive variables $w_{2n}(s)$ which depend on $N_w$. In the right panels, we plot the dual unitarity variables $X_J$: the area below those curves contribute to the dual objective $D^+$ and from the plot is clear that almost the whole contribution to the bound comes from the  spin-$0$ partial wave only.

We also checked numerically that dual unitarity $\det \Lambda^+_J(s)\approx 0$. According to the discussion in Appendix \ref{phapp}, this allows us to extract reliably the phase shifts using \eqref{dualphaseshifts}.
In figure \ref{dualps_max} we plot the real phase shifts for $J=0,2,4$ using the data obtained with $L=4$ as function of $N_w$. The green dashed lines represent the primal phase shifts obtained with $N_\text{max}=5$ and $L_\text{max}=10$.
The phase shifts for $J>0$ are all small and positive with positive scattering length (the slope at threshold), which is compatible with an extremal amplitude dominated by the spin 0 partial wave, where the higher spins 
are barely excited.

\subsection{Minimum coupling dual problem}
\begin{figure}[t]
\centering
        \includegraphics[scale=0.268 ]{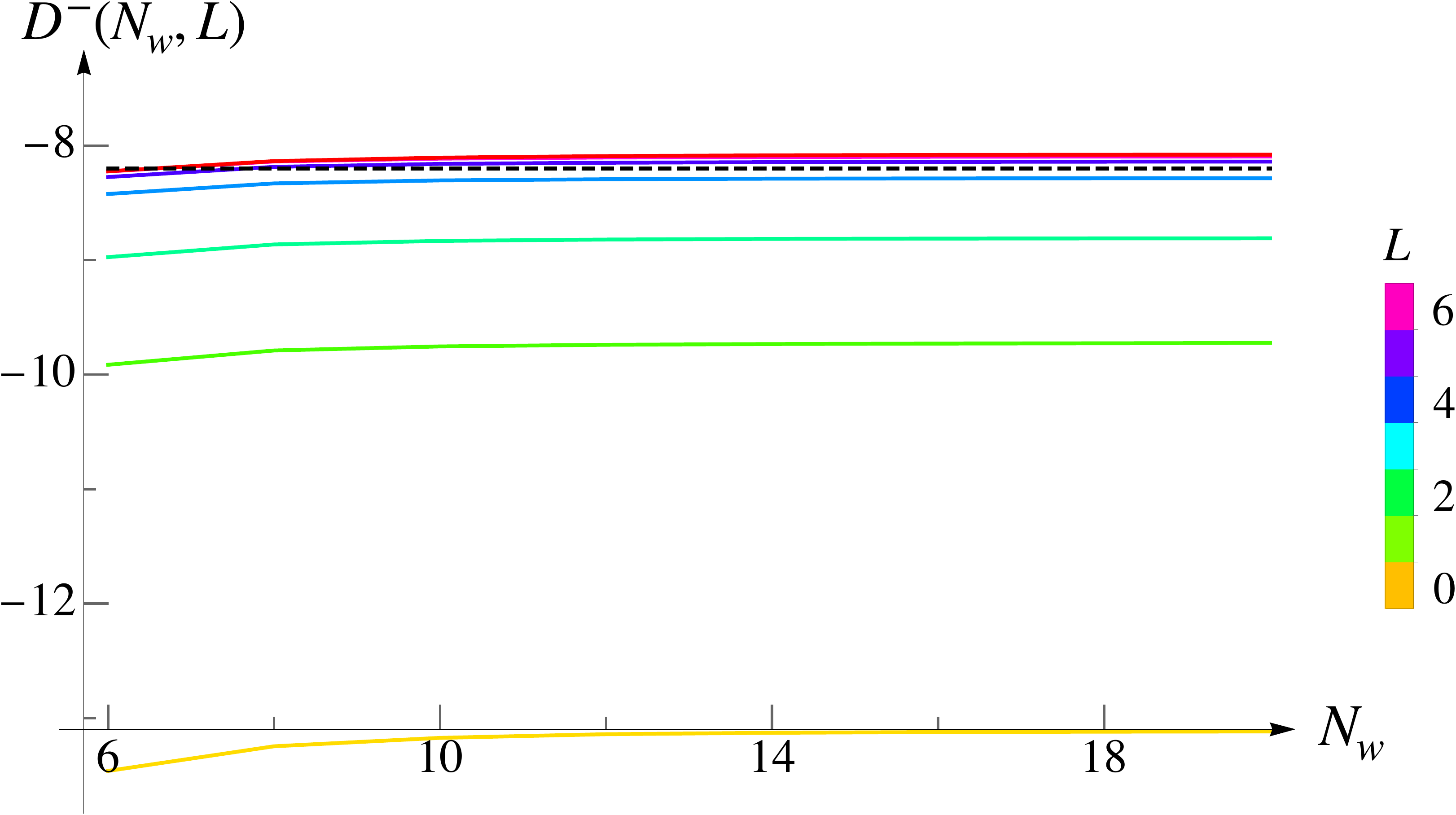}
\caption{Dual objective $D^-(4/3,4/3)$ for the minimum coupling problem as a function of $N_w$. Different colors correspond to different values of $L$. Here we see that the effect of the odd $w_{2n+1}$ is important for the convergence of the bound. The black dashed line represents the bound obtained in \cite{Lopez:1976zs}. Also in this case the bound we get improves the old results and is the best rigorous bound obtained so far: $g_0\geq -8.08$.} 
\label{min_bound}
\end{figure}

The minimum coupling problem is one of the hardest $S$-matrix Bootstrap problems since the corresponding optimal amplitude maximizes the total cross-section, see section \ref{numerical_results}. This difficulty turns into a much slower convergence of the dual problem in comparison to the one for the maximum coupling. In figure \ref{min_bound} we study the maximum of the dual-functional $D^-(4/3,4/3)$, denoted shortly as $D^-$, as a function of $N_w$ for different values of $L$. 

The dashed black line corresponds to the bound obtained by Lopez and Mennessier \cite{Lopez:1976zs} $g_0\geq -8.2$. Taking the bound obtained with $L=6$ and $N_w=20$ and extrapolating for $J_\text{max}\to\infty$ we can claim that $g_0\geq -8.08$. The difference $\Delta$ between the extrapolated value and the bound we get for the higher value of $J_\text{max}=26$ is $\Delta=4\times 10^{-3}$. The (small) dependence on $J_\text{max}$ means we still have some dual constraint violations. An inspection shows that the violations happen in the outer region $v\geq v_c$ in some intermediate range of energies that depend on the spin $J$. It would be interesting to estimate this region and prevent these violations by adding some fine-tuned constraints. However, as we said, their effect on the bound is very small.

Another difference w.r.t the maximum coupling problem is that every time we add an odd spin constraint the bound changes by a comparable amount to the one generated by adding the even ones, see figure \ref{min_bound}.

\begin{figure}[t]
\centering
        \includegraphics[scale=0.28 ]{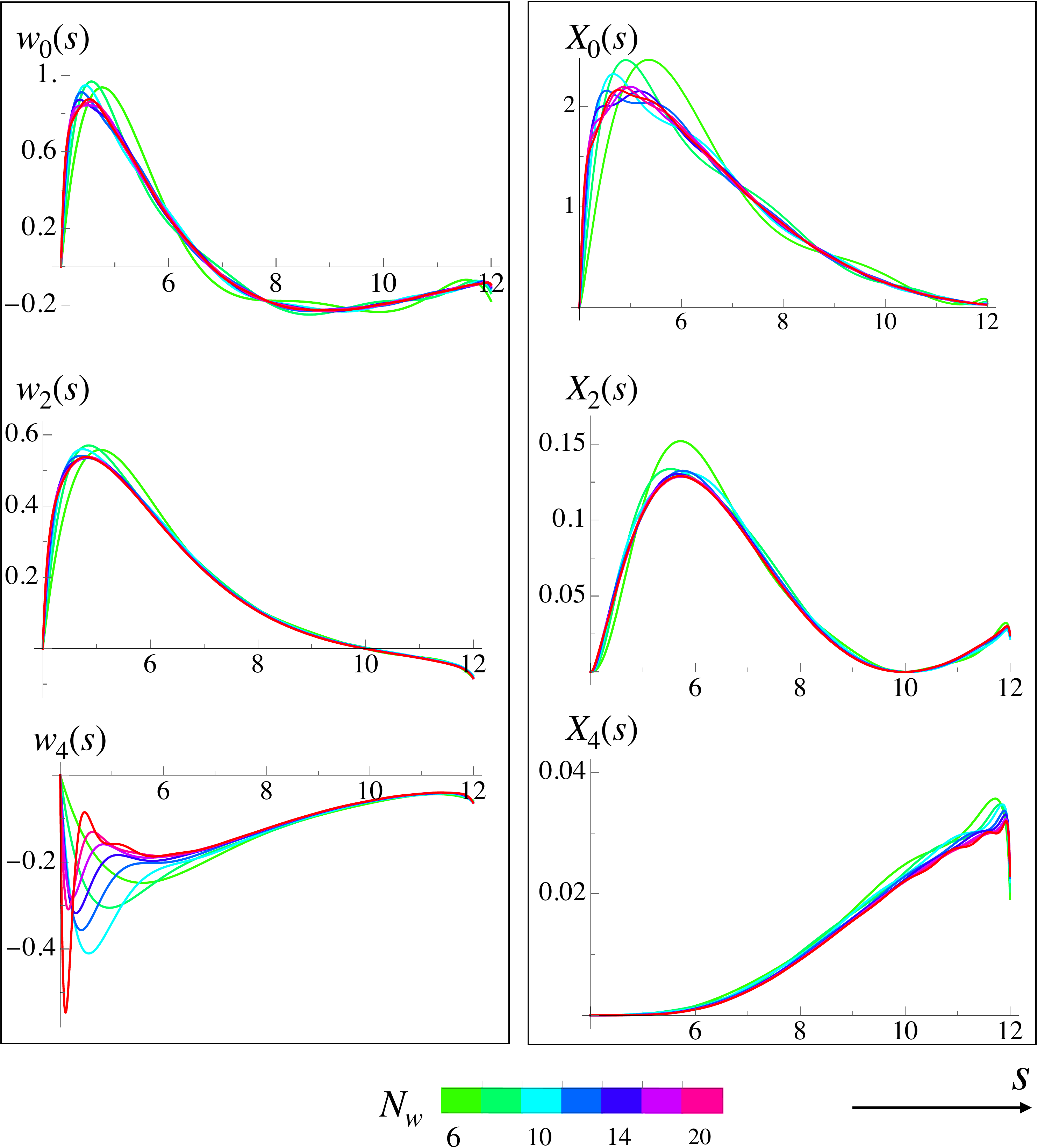}
\caption{Similarly to figure \ref{dualvars_max}, on the left panel we have the dual dispersion variables $w_{2n}$ for $n=0,1,2$, and on the right the unitarity dual variables. The dual objective $D^-$ is obtained by summing the areas below the $X_J$ curves with a negative sign. In this case, spin-2 and spin-4 contribute respectively to the $5\%$ and $1\%$ of the bound. Convergence in $N_w$ is also achieved for all dual variables, but $w_4$. We have observed that higher spins tend to stabilize as we increase $L$. Allowing for a more general threshold behavior might also help $w_4$ to stabilize. We leave this problem for the future as it does not affect the numerical stability of the bound, nor of the physical phase shifts. } 
\label{dualvars_min}
\end{figure}

In figure \ref{dualvars_min} we show the dual variables $w_{2n}$ (left panles) and $X_{2n}$ (right panels) up to $n=2$.
Again, it is worth noticing how the dual variables for this problem have a non-trivial higher spin structure compared to the maximum coupling case, figure \ref{dualvars_max}. 
The contribution to the dual objective coming from $X_2$ and $X_4$ accounts respectively for the $5\%$ and $1\%$ of the bound. Moreover, though the dual variable $w_4$ does not seem to converge yet in $N_w$, we observe that $X_4$ is stable. From our preliminary explorations it seems that as we add higher spin waves, the lower spins tend to stabilize. We expect that this will happen to the spin-4 dual variable if we were to increase further $L$. We leave this study to the future.

\begin{figure}[t]
\centering
        \includegraphics[scale=0.33 ]{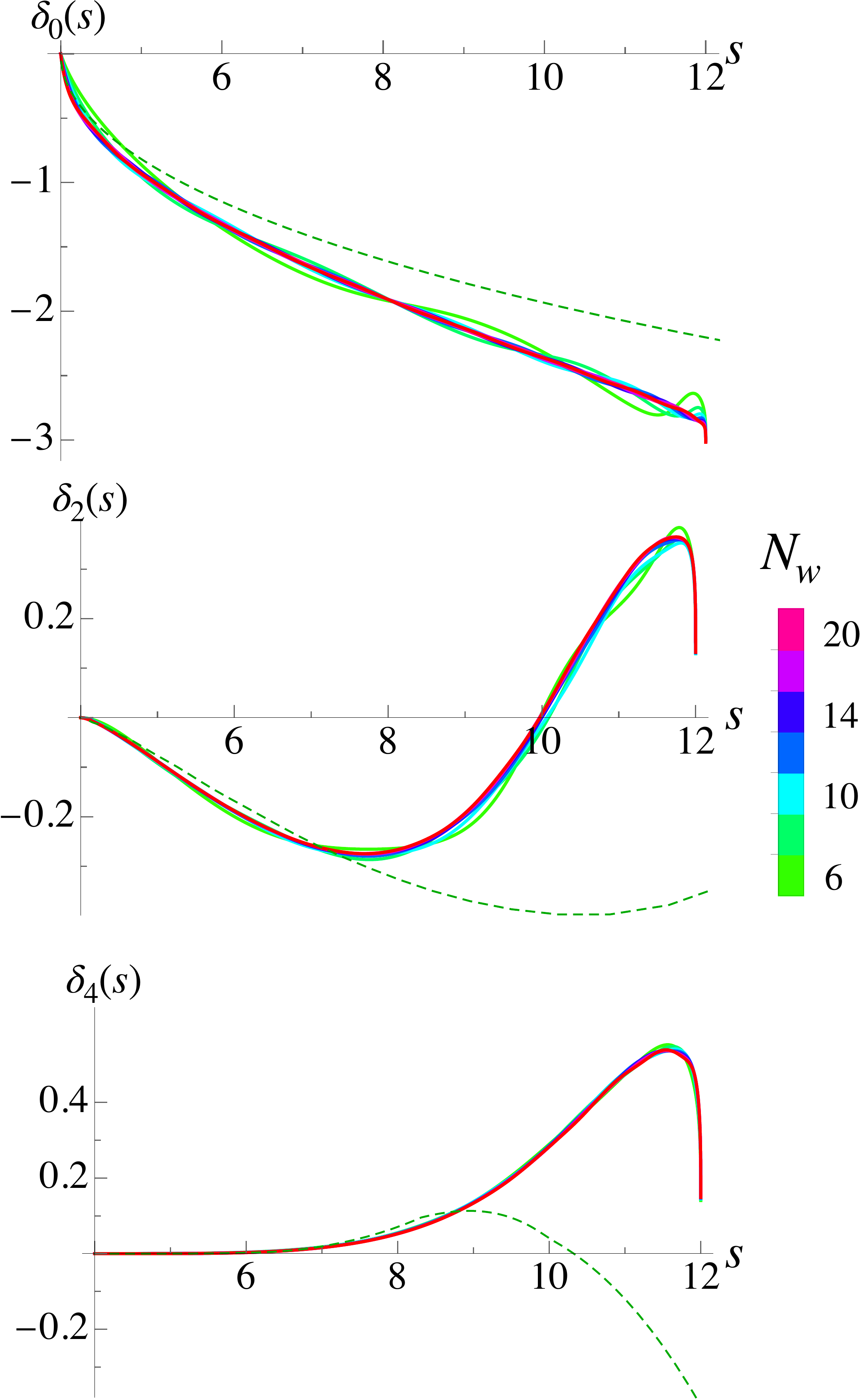}
\caption{Phase shifts $\delta_\ell(s)=\tfrac{1}{2i}\log{S_\ell(s)}$ for the first few spins as a function of $s$ of the amplitude that minimizes the quartic coupling. In green dashed we plot the primal phase shifts for our best numerics $N_\text{max}=20$, $L_\text{max}=22$. Solid lines in color gradient are obtained from the dual numerics with $L=4$ and different $N_w$. In this case, primal and dual differ significantly, especially for higher spins. The duality gap is also relatively big $\Delta_\text{gap}=2\tfrac{|g_0-D^-|}{|g_0+D^-|}=0.26$. One reason for this discrepancy comes from primal numerics as both the bound and unitarity saturation are still far from being attained. Nonetheless, the threshold physics is compatible, and we do expect they will agree more as the duality gap shrinks -- see also figure \ref{phaseshifts}.} 
\label{dualps_min}
\end{figure}

Finally in figure \ref{dualps_min} we compare primal (dashed-greed) and dual (color gradient) physical phase shifts for spin $\ell=0,2,4$.
In this case, the ratio between the duality gap and the bound is still large. Therefore, it does not come as a surprise that the phase shifts are different. This difference is more pronounced as we go to higher spins and energies. The spin-$0$ phase shift $\delta_0$ agrees nicely. In particular, this phase shift and the corresponding scattering length are negative for both primal and dual problems. For spin-2 and spin-4 the threshold behavior seems to agree, but at intermediate energies, they differ significantly. In this case, we observe that unitarity is not well saturated for primal both for spin-2 and spin-4. Hence, it might be useful to further improve the primal and achieve better convergence before drawing any conclusion.

As a last comment, it seems that the spin-2 scattering length for the minimum coupling problem is negative, while for spin-4 and higher becomes positive. 
This seems to be in contrast with the expectation coming from Froissart-Gribov representation that would suggest a positive spin-2 scattering length \cite{Yndurain:1972ix}.

\section{Threshold unitarity}
\label{th_unitarity}

In this appendix, we study the constraints that come from threshold unitarity in a $3+1$ dimensional gapped theory and make a connection to our numerical results.\footnote{See \cite{Correia:2020xtr} for a general analysis.}

We assume a threshold behavior of the form
\beqa
T(s,t,u)&=&-\alpha \left(\frac{1}{\sqrt{4-s}}+\frac{1}{\sqrt{4-t}}+\frac{1}{\sqrt{4-u}}\right)\\
&+&a_0+i b_0(\sqrt{s-4}+\sqrt{t -4}+\sqrt{u-4})\,.\nn
\eeqa
Its projection onto the spin-$0$ partial wave is
\beq
S_0(s)=1-\frac{\alpha}{32\pi}+i\sqrt{s-4}\frac{a_0-4b_0-\alpha}{32 \pi}+\mathcal{O}(s-4)\,.
\label{spin0_thresh}
\eeq

Expanding the elastic unitarity condition for $s\to 4^+$ implies that
\beq
\alpha=0\quad\text{or}\quad\alpha=64 \pi\,.
\eeq
and
\beq
b_0=(a_0-\alpha)/4\,.
\eeq

For $\alpha=64\pi$, the amplitude has a pole at threshold and the  spin-$0$ phase shift reads
\beq
\delta_0(s)=\pm \frac{\pi}{2}+\frac{1}{16}\left(\-2\mp\sqrt{100-\frac{a_0}{\pi}}\right)\sqrt{s-4}+\dots\,.
\eeq
It starts at $\pi/2$ and from its slope, we can extract the coefficient $a_0$.
In particular, the reality of the phase in the elastic unitarity region implies that $a_0\leq 100 \pi$. This is the behavior observed in figure \ref{phaseshifts} (top) and figure \ref{dualps_max} for the maximum coupling problem. 

For $\alpha=0$ we get
\beq
\delta_0(s)=\frac{1}{16}\left(-2\pm\sqrt{4+\frac{a_0}{\pi}}\right)\sqrt{s-4}+\dots\,,
\eeq
and the amplitude approach the constant $a_0$ at threshold, which is called \emph{scattering length}. This is the behavior observed in figure \ref{phaseshifts} (bottom), and figure \ref{dualps_min} for the minimum coupling.

The scattering length $a_0$ controls the time delay of the waves in the non-relativistic limit. A negative scattering length would signal a time advance. It is therefore expected to be bounded from below. Indeed, the reality of the phase now implies that $a_0\geq -4 \pi$.

\bibliography{dual_bootstrap} 

\end{document}